\documentclass[aps,prd,twocolumn,superscriptaddress,nofootinbib]{revtex4-2}

\usepackage{latexsym}
\usepackage{amsmath}
\usepackage{amssymb}
\usepackage{amsfonts}
\usepackage{bm}
\usepackage{physics}

\usepackage{color}
\definecolor{purple}{rgb}{0.5,0,0.5}
\definecolor{blue}{rgb}{0.0,0,0.9}
\definecolor{prdblue}{rgb}{0.133,0.118,0.498}
\usepackage[colorlinks=true, pdfstartview=FitV, linkcolor=prdblue, citecolor= prdblue, urlcolor=prdblue]{hyperref}

\usepackage{supertabular} 
\usepackage{placeins}
\usepackage{epsfig}
\usepackage{graphicx}

\begin{document}


\title{Sexaquarks and $H$ dibaryons in the $uuddss$ system: a comparison within a constituent quark model}


\author{M.C. Gordillo}
\email[]{cgorbar@upo.es}
\affiliation{Departamento de Sistemas F\'isicos, Qu\'imicos y Naturales, Universidad Pablo de Olavide, E-41089, Dos Hermanas, Sevilla, Spain}
\affiliation{Instituto Carlos I de Física Teórica y Computacional,
Universidad de Granada, E-18071 Granada, Spain.}


\date{\today}

\begin{abstract}

We study the $uuddss$ multiquark within a constituent quark model framework, solving the corresponding nonrelativistic Schr\"odinger equation including the AL1 potential by means of a diffusion Monte Carlo (DMC) method. 
The total wavefunction is written as the product of a radial component and an exact spin–color–flavor state, restricted to  isospin $I$=0.  
For this isospin, all allowed flavor wave functions are included. 
We explore two distinct constructions of the six–quark system. In the first one,  corresponding to a sexaquark, all six quarks are treated as indistinguishable and the wave function is fully antisymmetric with respect to the exchange of any two quarks.In the second one,  corresponding to the $H$ dibaryon, the system is partitioned into two sets of three quarks, effectively mimicking a baryon–baryon-like configuration including hidden color terms in which antisymmetry is imposed only within each three–quark cluster.
Only when the system is forced into a baryon–baryon–like configuration,  and for certain values of the spin, color and flavor quantum numbers, do we obtain states with masses close to, but above, the two-baryon threshold. Those states are characterized by two loosely bound three-quark clusters separated from one another by a distance of $\sim$ 2.5 fm.  The remaining structures are compact objects irrespectively of their internal wavefunction.  
\end{abstract}



\maketitle


\section{Introduction}

The possible existence of the $H$ dibaryon remains one of the long standing open problems in hadron physics. 
Originally proposed by Jaffe within the MIT bag model \cite{Jaffe1977}, the $H$ dibaryon was suggested to be a six-quark $I=0$, $J^P=0^+$ state with quark content $uuddss$ and a mass of about 2150 MeV, well below the $\Lambda\Lambda$ threshold. 
In the same work, another state with identical quark content, $I=0$, $J^P=1^+$, and a mass around 2335 MeV was predicted, potentially appearing as a structure in the $\Lambda\Lambda$ invariant-mass spectrum.  

Jaffe’s prediction triggered extensive experimental searches for the $H$ dibaryon, but none has confirmed a deeply bound state. The strongest constraints come from double-$\Lambda$ hypernuclei, which probe the $\Lambda\Lambda$ interaction through binding energies. In particular, the analysis of the NAGARA event \cite{NAGARA} indicates that, if the $H$  dibaryon exists, it cannot be deeply bound and should lie close to or above the $\Lambda\Lambda$ threshold.   Other experimental studies \cite{Yamamoto2000,PRC2007,PhysRevLett.110.222002,ALICE2016,ALICE2019}, including  recent results \cite{He2025},  are also consistent with a weakly bound or unbound scenario. This picture has been partly revisited in recent theoretical work \cite{Gal2024}, which suggests that a bound $H$ dibaryon may still be compatible with the NAGARA constraints.

On the other hand, the $H$ dibaryon has been extensively studied using a wide variety of theoretical approaches, starting with the original work by Jaffe \cite{Jaffe1977}, which predicts a deeply bound $H$ dibaryon. A similar picture emerges from most lattice QCD studies \cite{Beane2011,Inoue2011,Francis2019,Green2021,AokiDoi2020,
Sasaki2020}, although the binding is significantly reduced or even disappears as one approaches physical quark masses \cite{Shanahan2013}.
Other treatments produce similar outcomes \cite{Kodama1994,Haidenbauer2011,Carames2012}. One can summarize the theoretical consensus by saying that the $H$ dibaryon is either weakly bound or unbound. 
 
Most previous experimental and theoretical interpretations consider, either directly or indirectly,  the $uuddss$ multiquark system to be  an ensemble of two baryons,  typically a $\Lambda\Lambda$ configuration or a combination of 
$\Lambda\Lambda$, $N\Xi$, and $\Sigma\Sigma$ pairs, i.e., a $H$ dibaryon.  However,  it has been suggested that an alternative structure, referred to as a sexaquark \cite{Doser2023,Farrar2023,She2025}, may exist for the same system. This would correspond to a compact configuration 
in which all quarks are
treated as indistinguishable in the $SU(3)_f$ symmetric limit. The existence and properties of such a fully antisymmetric six-quark state remain largely unexplored, with no indication of its possible mass. On the other hand, it is known that this state cannot be reconstructed using naive baryon-baryon couplings alone \cite{Farrar2023}. 
 
 In this work we investigate the $uuddss$ system within a constituent quark model using the AL1 interaction \cite{SilvestreBrac1996}.  
The total wave function is written as the product of a radial component and an internal spin–color–flavor wave function,
\begin{equation}
\Psi = \Psi_{\mathrm{radial}} \, \Psi_{\mathrm{scf}} . 
\end{equation}
The same totally symmetric radial function is used for all multiquark configurations. 
From a given set of spin–color–flavor functions, a baryon–baryon-like arrangement is obtained by imposing antisymmetry restrictions on the first three and the last three quarks, while treating quarks belonging to different “baryons” as distinguishable. In this way, although all quarks are fermions, the Pauli principle is satisfied, since it applies only to identical (indistinguishable) quarks. This subset of the Hilbert space 
corresponds to configurations associated with the  $H$ dibaryon.
On the other hand,  for the {\em same} set of spin-color-flavor functions, a compact structure is obtained when all quarks are treated as equivalent and antisymmetry is imposed under the exchange of all six quarks. This subset of the Hilbert space describes the sexaquark. 
This procedure, in which different structures are explored by imposing different antisymmetry conditions on the same set of quarks, has already been used in the literature \cite{Gordillo2024,Gordillo2025} and allows for a direct comparison between different configurations within the same framework. 
We find that those two configurations, $H$ dibaryon and sexaquark,  lead to significantly different masses, and therefore correspond to distinct states that could, in principle, be observed experimentally.

The manuscript is organized as follows:
in Sec. 2 we describe the constituent quark Hamiltonian and
briefly outline the DMC formalism.  Section 3 presents our numerical results for both the $H$ dibaryon and the sexaquark,  including masses and structural parameters.  Finally, Sec. 4 contains our conclusions.
 
\section{Method}
To describe a set of quarks within the non-relativistic quark model, we have to solve the Schr\"odinger equation defined by  the corresponding Hamiltonian.  That Hamiltonian includes the AL1 empirical potential from Ref. ~ \cite{SilvestreBrac1996}, and for a set $n$ particles reads:
\begin{equation}
H = \sum_{i=1}^n\left(m_i+\frac{\vec{p}^{\,2}_i}{2m_i}\right) - T_{\text{CM}} + \sum_{i<j} V(\vec{r}_{ij}) \,,
\label{eq:Hamiltonian}
\end{equation}
where $m_{i}$ is the quark mass, $\vec{p}_i$ its momentum and $T_{\text{CM}}$ the center-of-mass kinetic energy. 
$V(\vec{r}_{ij})$ is a quark-quark interacting potential that contains both one-gluon exchange and color confining interactions:
\begin{equation}
V(\vec{r}_{ij}) = V_{\text{OGE}}(\vec{r}_{ij}) + V_{\text{CON}}(\vec{r}_{ij})\,,
\end{equation}
with 
\begin{align}
V_{\text{OGE}}(\vec{r}_{ij}) = \frac{1}{4} \alpha_{s} (\vec{\lambda}_{i}\cdot
\vec{\lambda}_{j}) \Bigg[\frac{1}{r_{ij}} 
- \frac{2\pi}{3m_{i}m_{j}} \delta^{(3)}(\vec{r}_{ij}) (\vec{\sigma}_{i}\cdot\vec{\sigma}_{j}) \Bigg] \,,
\label{eq:OGE}
\end{align}
and
\begin{equation}
V_{\text{CON}}(\vec{r}_{ij}) = (b\, r_{ij} + \Delta) (\vec{\lambda}_{i}\cdot
\vec{\lambda}_{j}) \,,
\label{eq:CON}
\end{equation}
where  $\vec{\lambda}$ and $\vec{\sigma}$ are,  respectively,  the  Gell-Mann  color and  Pauli spin matrices.  $\delta^{(3)}(\vec{r}_{ij})$ is replaced by:
\begin{equation}
\delta^{(3)}(\vec{r}_{ij}) \to \kappa \, \frac{e^{-r_{ij}^2/r_0^2}}{\pi^{3/2}r_{0}^3} \,,
\end{equation}
where $r_0 = A \left(\frac{2m_im_j}{m_i+m_j}\right)^ {-B}$. 
All the parameters needed to define the potential are given in 
Table~\ref{tab:parameters}.  

\begin{table}[!t]
\caption{\label{tab:parameters} AL1 potential parameters as defined in  Ref.~\cite{SilvestreBrac1996}.}
\begin{ruledtabular}
\begin{tabular}{llc}
Quark masses & $m_c$ (GeV) & 1.836 \\
             & $m_b$ (GeV) & 5.227 \\
\hline
OGE          & $\alpha_s$        & 0.3802 \\
             & $\kappa$          & 3.6711 \\
             & $A$ (GeV)$^{B-1}$ & 1.6553 \\
             & $B$               & 0.2204 \\
\hline
CON          & $b$ (GeV$^2$)  &  0.1653 \\
             & $\Delta$ (GeV) & -0.8321 \\
\end{tabular}
\end{ruledtabular}
\end{table}
To solve the Schr\"odinger equation derived from the Hamiltonian in Eq. \ref{eq:Hamiltonian}, we used the diffusion Monte Carlo (DMC) algorithm.  We start by writing the Schr\"odinger equation in imaginary time ($\hbar=c=1$) as \cite{hammond}:
\begin{equation}
-\frac{\partial \Psi (\bm{R},\alpha,t)}{\partial t}
= (H-E_s)\,\Psi(\bm{R},\alpha,t)\,,
\label{eq:Sch1}
\end{equation}
where $E_s$ is an appropriate energy shift and $\bm{R}\equiv(\vec{r}_1,\ldots,\vec{r}_n)$ stands for the coordinates of the $n$ quarks.  $\alpha$ indicates each possible spin-flavor-color state (channel in the standard literature) 
with fixed quantum numbers.  DMC is then an operational recipe 
to obtain the ground state of the system.  It needs 
 an initial approximation 
to that ground state,  the so-called trial function, $\Psi_{T} (\bm{R}, \alpha)$, that includes 
all the information known a priori about the system.
In our approach,  the trial wave function is constructed
as,
\begin{align} \label{eq:Eigenfunction}
\Psi_{T}(\bm{R},\alpha)=\phi_T(\bm{R})\sum_\alpha d_\alpha\chi^{(\alpha)},
\end{align}
where $\phi_T(\bm{R})$ depends on the quark coordinates and the coefficients $d_\alpha$
encode the spin ($s_i$)-color ($c_i$)–flavor ($f_i$)  structure of the spin-color-flavor function $\chi^{(\alpha)} (s_i, f_i, c_i)$ corresponding to 
the $\alpha$ channel for  every $i$ quark.
In the DMC algorithm with importance sampling,  the evolution of the system is guided not by $[\Psi_{T}(\bm{R},\alpha)]^2$ as in a purely variational method,  but by the product $f(\bm{R},\alpha,t)$ =$\Psi (\bm{R},\alpha,t)  \Psi_{T} (\bm{R},\alpha)$.  This $f(\bm{R},\alpha, t)$ is represented by a set of {\em walkers} that include both the  particle coordinates {\em and} the coefficients of the spin-color-flavor functions.  In a DMC step,  both the coordinates and coefficients 
are sampled at the same time,  since the  Hamiltonian couple all channels.  A full description, together with the corresponding tests of the robustness of the method, can be 
found in Ref.  ~\cite{Gordillo2020}. 

In Eq.~\eqref{eq:Eigenfunction}, the $\chi^{(\alpha)}$ are constructed from the eigenfunctions of the spin, {\em flavor}, and color operators, defined as  
\begin{align}
S^2 &= \left(\sum_{i=1}^{6} \frac{\vec{\sigma}_i}{2}\right)^2 \,, 
F^2 &= \left(\sum_{i=1}^{6} \frac{\vec{\lambda}_i^f}{2}\right)^2 \,, 
C^2 &= \left(\sum_{i=1}^{6} \frac{\vec{\lambda}_i}{2}\right)^2 \,,
\label{eq:operators_penta}
\end{align}
with well-defined eigenvalues. Here, $\vec{\lambda}$ denotes the standard Gell-Mann matrices acting in color space, while $\vec{\lambda}_i^f$ are the same matrices acting in flavor space ($u$, $d$, and $s$ instead of the color labels $r$, $g$, and $b$).  
To obtain the flavor eigenfunctions,  we consider the full $SU(3)_f$ symmetry, i.e., we consider $u$, $d$ and $s$ as equivalent.  

We are, of course, restricted to the subspace of color-singlet states, whereas the flavor eigenvalues are not subject to such a constraint. However, since the quantum number typically measured in experiments is the isospin, the flavor functions must be constructed as {\em simultaneous} eigenvectors of both the isospin and flavor operators. The isospin operator is defined as  
\begin{equation}
I^2 = \left(\sum_{i=1}^{6} \frac{\vec{\lambda}_i^f}{2} \right)^2,
\end{equation}
and involves only the $\lambda^{f,1}_i$, $\lambda^{f,2}_i$, and $\lambda^{f,3}_i$ components of $\vec{\lambda}_i^f$. It is isomorphic to the spin operator in the two-flavor ($u$, $d$) sector.  
We obtain the eigenvectors of $I^2$ and determine the eigenfunctions of $F^2$ within that restricted space \cite{Gordillo2025}. In this work, we focus on the $I^2=0$ sector, as in previous studies on the $uuddss$ system \cite{Farrar2023},  that is an subspace of dimension 30.  The first two of that set of 30 functions are:
\begin{equation}
\chi_{f,1}^{(12)} \;=\; \frac{1}{2}\Big(
|ssuudd\rangle - |ssuddu\rangle - |ssduud\rangle + |ssdduu\rangle
\Big)\,\nonumber
\end{equation}
\begin{align} 
\chi_{f,2}^{(12)} \;=\; \frac{1}{2\sqrt{3}}\Big(
|ssuudd\rangle - 2|ssudud\rangle +
|ssuddu\rangle \nonumber \\
+ |ssduud\rangle - 2|ssdudu\rangle + |ssdduu\rangle
\Big)\,,
\label{isospin}
\end{align}
where the two $s$ quarks occupy positions $1$ and $2$, as indicated by the superscript. The remaining 28 functions are obtained by keeping the relative positions of the $u$ and $d$ quarks fixed while varying the positions of the $s$ quarks. For example,  
\begin{equation}
\chi_{f,1}^{(24)} \;=\; \frac{1}{2}\Big(
|ususdd\rangle - |usdsdu\rangle - |dsusud\rangle + |dsdsuu\rangle
\Big)\, \nonumber 
\end{equation}
\begin{align}
\chi_{f,2}^{(24)} \;=\; \frac{1}{2\sqrt{3}}\Big(
|ususdd\rangle - 2|usdsud\rangle +
|usdsdu\rangle \nonumber \\
+ |dsusud\rangle - 2|dsusdu\rangle + |dsdsuu\rangle
\Big)\,,
\end{align}
correspond to the case where the $s$ quarks occupy positions $2$ and $4$. From these 30 functions, we construct three sets of eigenvectors of the flavor operator with different eigenvalues: 5 functions with $F^2=0$, 16 with $F^2=3$, and 9 with $F^2=8$. These functions are listed in Appendix~I. We combine them with the 5 colorless functions and the appropriate number of spin eigenvectors (5 for $S=0$, 9 for $S=1$, 5 for $S=2$, and 1 for $S=3$) to construct candidates for the $\chi^{(\alpha)}$ channels in Eq.~\ref{eq:Eigenfunction}. The 5 colorless functions are isomorphic to the flavor ones in the $F^2=0$ case under the mapping $u \leftrightarrow r$, $d \leftrightarrow g$, and $s \leftrightarrow b$.The eigenvectors of the spin operator for the four spin values considered in this work are given in Appendix~II. 

At this point, a clarification is in order. In many cases, several eigenfunctions correspond to the same eigenvalue, so any linear combination of them is also a valid choice in constructing a $\chi^{(\alpha)}$ channel. 
For instance, it is not necessary to use the 5 $\zeta_f$ functions listed in Appendix~I to describe the $F^2=0$ sector; any set of 5 functions spanning the same flavor subspace would be equally suitable. The reason is that physical observables depend only on the subspace itself, not on a particular choice of basis.  The same can be said of the spin and color degrees of freedom.  The dimension of the corresponding Hilbert subspaces for the different values of flavor and spin (with $I^2$ =0 in all cases) are given in Table \ref{functions}.

\begin{table}[h]
\caption{Number of functions,$N_s$, for each value of spin, $S$, together with the number of functions, $N_f$, for each eigenvalue of the flavor operator, $F^2$.  The total number of spin-color-flavor combinations, $N_T$, includes the multiplicity due to the 5 colorless states.  We also show the dimension of the spin-color-flavor space with the proper antisymmetry restrains for the sexaquark and the $H$ dibaryon. $-$ stands for the absence of functions with the proper symmetry. } 
\label{functions} 
\centering
\begin{tabular}{c c c c c c c}
\hline
$S$ & $N_s$ & $F^2$ & $N_f$ & $N_T$ & Sexaquark & $H$ dibaryon  \\
\hline
0 & 5 & 0 & 5 &  125 &  1 & 4\\
0 & 5 & 3 & 16 &  400 & - & 10 \\
0 & 5 & 8 & 9 &  225 & 1 & 7\\
\hline
1 & 9 & 0 & 5 & 225 & --  & 6 \\
1 & 9 & 3 & 16 & 720 & 1  & 20 \\
1 & 9 & 8 & 9 & 405 &  -- & 13 \\
\hline
2 & 5 & 0 & 5 & 125 & -- & 3 \\
2 & 5 & 3 & 16 & 400 &  1 & 12\\
2 & 5 & 8 & 9 &  225 &  1 & 8 \\
\hline
3 & 1 & 0 & 5 &  25 & -- & 1 \\
3 & 1 & 3 & 16 &  80 & -- & 2 \\
3 & 1 & 8 & 9 & 45 & -- & 2 \\
\hline
\end{tabular}
\end{table}

Now, we have to take each set of functions and impose antisymmetry on the exchange of quarks, since quarks are fermions. 
This is done in the spin-color-part of the wavefunction, since the radial part is totally symmetric due to the constraints applied to it (see below). A wavefunction is antisymmetric when it
is an eigenvector of the operator:
 \begin{equation}
\mathcal{A} = \frac{1}{N_p} \sum_{{\alpha}=1}^{N_p} (-1)^P \mathcal{P_{\alpha}} \,.
\label{eq:antisymope}
\end{equation}
Here, $N_p$ is the number of possible permutations of the quark indexes, $P$ denotes the order of the permutation, and the $\mathcal{P}_{\alpha}$ are the corresponding permutation matrices.  
Once the matrix associated with the operator in Eq.~\eqref{eq:antisymope} is constructed, using as a basis one of the 12 sets of spin-color-flavor functions defined above, we obtain its eigenvectors as linear combinations of the spin-color-flavor functions within each set. Only those combinations with eigenvalue equal to one are antisymmetric under the exchange of quarks.  
Different sets of antisymmetric functions can be obtained depending on the choice of $\mathcal{P}_{\alpha}$ in Eq.~\eqref{eq:antisymope}, since these determine which quarks are considered identical and therefore indistinguishable. If all six quarks are taken to be identical, the number of permutations exchanging any pair is $N_p = 720$. The resulting eigenvectors are wavefunctions fully antisymmetric under the exchange of {\em any} quark, corresponding to a sexaquark configuration.  
On the other hand, if only the first three quarks (and separately the last three) are identical, while being distinguishable from the other group, the number of admissible permutations is reduced to $N_p = 36$. This subset of the spin-color-flavor space corresponds to the $H$ dibaryon.  

The number of functions, or equivalently, the dimension of the subspace spanned by the antisymmetric states for each set of allowed quantum numbers, is listed in Table~\ref{functions}. These functions define the channels, $\chi^{(\alpha)}$, that enter Eq.~\eqref{eq:Eigenfunction} and are used to construct the trial wave function in the DMC method.  For instance, for the case with the quantum numbers $C^2$=0, $S$=0 and $F^2$=0, we have a single possible channel for the sexaquark and four  $\chi^{(\alpha)}$ channels for the $H$ dibaryon.  In this case,  those four channels are combined to describe the system via Eq. \ref{eq:Eigenfunction}, so we have a coupled channels calculation.  Again, it is important to stress that to generate this functions we consider $u$, $d$ and $s$ to be completely equivalent: at this stage, we do not single out the $s$ quark in any way.

Finally, to define completely the definition of the trial function given in Eq. \ref{eq:Eigenfunction}, we have to give a expression for 
$\phi_T(\bm{R})$.  In this work, we restrict ourselves to total orbital angular momentum $L^2=0$. This implies that the function should depend only on the set of inter-particle distances. This makes it symmetric under permutations of the spatial coordinates.  As in previous works \cite{Gordillo2020,Gordillo2024,Gordillo2025,Alcaraz-Pelegrina2022,Gordillo2023,Gordillo2024b}, we used the expression:  
\begin{equation}
\phi_T (\vec{r}_1,\ldots,\vec{r}_n) = \prod_{i<j} \exp(-a_{ij} r_{ij}) \,,
\label{eq:radialwf}
\end{equation}
where the coefficients $a_{ij}$ are chosen so as to satisfy the cusp conditions derived from the Coulomb-like part of the AL1 potential for each pair of constituents. 
We do not consider spin-orbit or tensor interactions because they are not included in the Hamiltonian defined in Eq. \ref{eq:Hamiltonian}.
It is nevertheless useful to clarify the consequences of
such terms within the spatial model space considered here.

The spatial function in Eq.~\ref{eq:radialwf} is real and depends only on interparticle
distances. Let us consider a generic pairwise spin-orbit interaction
\begin{equation}
 V^{LS}_{ij}
 =
 v_{ij}(r_{ij})\,\mathbf l_{ij}\cdot\mathbf s_{ij}.
\end{equation}
For arbitrary constituent masses $m_i$ and $m_j$, the relative orbital
angular momentum can be written as
\begin{equation}
 \mathbf l_{ij}
 =
 -i\hbar\,\mathbf r_{ij}\times D_{ij},
 \qquad
 D_{ij}
 =
 \frac{m_j}{m_i+m_j}\nabla_i
 -
 \frac{m_i}{m_i+m_j}\nabla_j .
\end{equation}
The spatial matrix element is therefore, after factoring out the spin-dependent part that depends only on the spin-color-flavor part of the total wavefuncion: 
\begin{equation}
 \mathbf A_{ij}
 =
 \left\langle\phi_T\left|
 v_{ij}(r_{ij})\,\mathbf l_{ij}
 \right|\phi_T\right\rangle
 =
 -\frac{i\hbar}{2}
 \int d\mathbf R\,
 v_{ij}(r_{ij})\,
 \mathbf r_{ij}\times D_{ij}\phi_T^2 ,
\end{equation}
where we have used the fact that $\phi_T$ is real. Integrating by parts,
and using
\begin{equation}
 D_{ij,c}r_{ij,b}=\delta_{cb},
 \qquad
 D_{ij,c}r_{ij}=\frac{r_{ij,c}}{r_{ij}},
\end{equation}
gives
\begin{equation}
 (A_{ij})_a
 =
 \frac{i\hbar}{2}
 \int d\mathbf R\,\phi_T^2
 \left[
 \frac{v'_{ij}(r_{ij})}{r_{ij}}
 \epsilon_{abc}r_{ij,b}r_{ij,c}
 +
 v_{ij}(r_{ij})\epsilon_{abc}\delta_{bc}
 \right]
 =0,
\end{equation}
since both contractions with the antisymmetric Levi-Civita tensor vanish.
Thus,
\begin{equation}
 \left\langle
 v_{ij}(r_{ij})\,\mathbf l_{ij}\cdot\mathbf s_{ij}
 \right\rangle=0
\end{equation}
for every pair $ij$, independently of whether the constituent masses are
equal or different. Hence, a radial pairwise spin-orbit interaction gives
no contribution within the the pure $L$=0 spatial sector considered
here. 
It is also worth noting that this conclusion is consistent with the only
explicit spin-orbit extension of the AL1 interaction that we have found in
the literature,  developed for a different system
\cite{Lemaire2002}. Although the quark--quark interaction employed in that
work contains a phenomenological pairwise spin-orbit term, the authors
explicitly state that, for the $L=0$ wave, there is no contribution from the
spin-orbit kernels, so that only the central and hyperfine interactions
contribute.

Similarly,  the matrix elements of a tensor interaction vanishes
within a pure $L=0$ sector. A tensor force produces nonzero effects only through coupling to partial waves with $L\neq 0$. Such orbital admixtures are not included in the
present calculation, since their inclusion would take the wave function
outside the pure $L=0$ Hilbert subspace assumed from the outset.

After completing this construction, the DMC algorithm can be applied to the 12 different trial functions, each of them with a fixed set of quantum numbers,  that we obtain by multiplying the linear combination of the different channels, $\chi^{(\alpha)}$, by Eq. \ref{eq:radialwf}.   As indicated above, a {\em walker} is a set of particle coordinates {\em and} coefficients $d_{\alpha}$ that determine the spin-color-flavor component of the wave function.  Those components include 
combination of terms corresponding to different flavor permutations of the light quarks, as can be seen, for instance,  in \ref{isospin}.  This means that we have to evaluate the quantities that depend  explicitly on the quark masses, such as the kinetic and potential energies as weighted averages over the internal states defined by the coefficients $d_{\alpha}$ \cite{Gordillo2025}.

The DMC technique is affected by a time-step error arising from the
application of the short-time approximation \cite{hammond} to the exact
Green's function of the system. Therefore, the imaginary-time step must be
small enough for this approximation to remain reliable, but also large
enough to ensure an adequate sampling of the relevant configuration space.
We have checked that a time step of $10^{-6}$ MeV$^{-1}$ satisfies both
requirements. This value has also been used in previous calculations
\cite{Gordillo2020, Alcaraz-Pelegrina2022, Gordillo2023, Gordillo2024,
Gordillo2025}. All the quantities reported here are averages over
simulation runs of $5\times 10^5$ steps, after discarding the first
$10^5$ steps to ensure equilibration. No appreciable drifts were observed
during the production runs.
To reduce spurious correlations within each simulation, only configurations
separated by 500 steps were included in the averages, yielding a sample of
1000 points. The corresponding statistical uncertainties are reported
together with the mean values in the tables. To check the robustness of the
results, each simulation was repeated at least three times, starting from
different initial configurations, and the resulting averages were found to
be consistent within the quoted error bars.
Finally, all simulations were performed using $10^3$ walkers. Increasing
the walker population did not lead to any noticeable improvement in the
quality of the results.

\section{Results}

\subsection{Masses}

In this subsection, we present the  DMC results for the masses for  the sexaquark and the $H$ dibaryon.  In all cases, the isospin was fixed to $I^2$ =0, and we considered all the eigenvalues of the flavor operator compatible with that restriction.  Since the AL1 potential employed in this work is empirical and we are using a non-relativistic approach, to know whether a $uuddss$ combination 
is stable with respect to the breaking up  into two baryons, we need to know the masses of the corresponding  baryons calculated within the same approximations.  Those masses are given in Table \ref{baryons}.  In this work, we used the AL1 potential with the parameters provided in Ref. \cite{SilvestreBrac1996}. In that work, the baryon masses derived from the Hamiltonian are corrected afterwards by a three-body term depending on the quark masses introduced ad hoc.  We have chosen to ignore that term since no recipe to apply it to multiquarks other than baryons was given. 

\begin{table}[h]
\caption{Masses, in MeV,  together with their error bars corresponding to the different baryons we use to calculate the
different stability  thresholds of the hexaquarks considered in this work. Also given the experimental masses from Ref. \cite{pdg}. The error bars are in the last figure. Since AL1 does not distinguished between $u$ and $d$ quarks, equivalent baryons are displayed together. The masses obtained by the DMC do not include the three-body term that corrects them in Ref. \cite{SilvestreBrac1996}}.  
\label{baryons}
\centering
\begin{tabular}{c c c c c c}
\hline
Baryon & $S$ & $F^2$ & $I$ & Mass & Mass$_{exp}$  \\
\hline
$p/n$ & 1/2 & 3 & 1/2 & 1030 & 938/940 \\
$\Delta^+/ \Delta^0$ & 3/2 & 6 & 3/2 & 1307 & 1232 \\ 
\hline
$\Lambda^0$ & 1/2 & 3 & 0 & 1217 &    1116 \\
$\Sigma^+/\Sigma^0/\Sigma^-$& 1/2 & 3 & 1 & 1217 & 1189/1193/1197 \\
$\Sigma^{*+}/\Sigma^{*0}/\Sigma^{*-}$& 3/2 & 6 & 1 & 1450 & 1383/1384/1387 \\
\hline
$\Xi^0/\Xi^-$ & 1/2 & 3 & 1/2 & 1380 & 1315/1322 \\
$\Xi^{*0}/\Xi^{*-}$ & 3/2 & 6 & 1/2 & 1574 & 1532/1535 \\
\hline
\end{tabular}
\end{table}

In Table \ref{masses} we give the values of the masses obtained for different  Sexaquarks and $H$ dibaryons obtained by  the DMC procedure described above,  that considers the properly constructed spin-color-flavor functions.  In that Table, we show also the minimum mass threshold corresponding to the pair of baryons compatible with the quantum numbers of the different hexaquarks.  Those thresholds were obtained using the data in Table \ref{baryons}.  

\begin{table}[t]
\caption{Masses (in MeV) with error bars for the sexaquark and $H$ dibaryon states for different $(S,F^2)$ sectors. The lowest compatible two-baryon thresholds, T,  and their masses are also shown. The mass differences between the sexaquark and $H$ dibaryon and their their thresholds are displayed for clarity. }
\label{masses}
\centering
\begin{tabular}{c c c c c c c c}
\hline
$S$ & $F^2$ & Sexaquark & $H$ dibaryon & T ($I=0$) & Mass &$\Delta$ M$_S$ & $\Delta$M$_H$  \\
\hline
0 & 0 & $2659\pm6$ & $2656\pm5$ & $N\Xi$ & 2410 & 249 &246  \\
0 & 3 & -- & $2446\pm4$ & $N\Xi$ & 2410 & -- &  36 \\
0 & 8 & $3068\pm4$ & $2436\pm6$ & $N\Xi$ & 2410 & 658 & 26 \\
\hline
1 & 0 & -- & $2564\pm5$ & $N\Xi$ & 2410 & -- & 154 \\
1 & 3 & $2871\pm7$ & $2553\pm4$ & $N\Xi$ & 2410 & 461 & 143 \\
1 & 8 & -- & $2448\pm3$ & $N\Xi$ & 2410 & --&  38 \\
\hline
2 & 0 & -- & $2933\pm6$ & $N\Xi^*$ & 2604 & -- & 329 \\
2 & 3 & $2934\pm4$ & $2899\pm6$ & $N\Xi^*$ & 2604& 330 & 295\\
2 & 8 & $3158\pm5$ & $2633\pm3$ & $N\Xi^*$ & 2604 & 554 & 29 \\
\hline
3 & 0 & -- & $3143\pm5$ & $\Sigma^*\Sigma^*$ & 2900 & -- & 243 \\
3 & 3 & -- & $3156\pm3$ & $\Sigma^*\Sigma^*$ & 2900 & -- & 256 \\
3 & 8 & -- & $3134\pm4$ & $\Sigma^*\Sigma^*$ & 2900 & -- & 234 \\
\hline
\end{tabular}
\end{table}

The results displayed in Table~\ref{masses} show a consistent pattern for all $S$ and $F^2$ values. First,  all calculated masses lie above the lowest compatible two-baryon thresholds. For $S=0$ and $S=1$, the relevant threshold is $N\Xi$ at 2410 MeV, while for higher spins the thresholds involve other  baryons and increase accordingly.  There are also above the $\Lambda \Lambda$ (2424 MeV) in the cases in which this limit is appropriate.  This indicates that, within this approximation,  neither the sexaquark configurations nor the $H$ dibaryon have stable bound states.  

Secondly,  a comparison between compact sexaquark configurations and $H$ dibaryon states, when the former are possible,  shows that $H$ dibaryons have  consistently lower masses, specially in the case of highest $F^2$ values. 
This can be explained by the larger number of channels available to the $H$ dibaryon configurations due to the symmetry constraints removed at constructing the internal wavefunctions (see Table \ref{functions}). 
This suggests that fully compact six-quark configurations are disfavored relative to baryon--baryon-like structures, even though sometimes not by much, as in the $S=0$, $F^2$ =0 case. 
We can see also that only a limited subset of $H$ dibaryons lie 
close to the thresholds.  Those are the corresponding to the 
$(S,F^2)=(0,3)$, $(0,8)$,  $(1,8)$ and $(2,8)$ sectors, which appear approximately $20$--$40$ MeV above the lowest threshold. 
These states can be interpreted as weakly unbound configurations, potentially sensitive to channel coupling effects or moderate changes in the interaction. Surprisingly, they correspond to excited states of the flavor operator, not to its ground state, $F^2$=0. 

Up to this point, we have considered wavefunctions with full $SU_f(3)$ symmetry, i.e., the $s$ quark is not singled out in any way. 
The difference between the masses of $u,d$ and $s$ quarks was
taking into account only in the DMC procedure.  
This implies, for instance, that in the $H$ dibaryon symmetry space one can have $\Lambda\Lambda$, $N\Xi$, and $\Sigma\Sigma$-{\em like}  combinations on an equal footing within the same $\chi^{(\alpha)}$.
However, we have to stress that this does {\em not} mean that we are considering direct products of baryon wavefunctions, since in each spin--color--flavor channel we include hidden-color components that are not present in a baryon--baryon description.
With this in mind, one can also consider internal wavefunctions in which the $s$ quark is treated as distinguishable. 
In this case, the resulting states are no longer eigenvectors of the flavor operator, but only of isospin. For instance, for $S=0$, if we fix the positions of the two $s$ quarks to be $1$ and $2$, only the two isospin functions given in \ref{isospin} are allowed. We can then repeat the DMC calculation in order to compare the effects of that symmetry breaking in the masses of the baryons and hexaquarks.
In that calculation each quark in a walker carries fixed labels corresponding to its mass and position.  
Those are collected in Tables \ref{barsin} and \ref{sin}, respectively. 

\begin{table}[h]
\caption{Same as in Table \ref{baryons} but for internal functions with flavor symmetry broken. No three-body correction is applied to the masses. }
\label{barsin}
\centering
\begin{tabular}{c c c c c}
\hline
Baryon & $S$ & $I$ & Mass & Mass$_{exp}$\\
\hline
$p/n$ & 1/2 & 1/2 & 1007 & 938/940 \\
$\Delta^+/ \Delta^0$ & 3/2  & 3/2 & 1308 & 1232 \\ 
\hline
$\Lambda^0$ & 1/2  & 0 & 1156 &    1116 \\
$\Sigma^+/\Sigma^0/\Sigma^-$& 1/2 & 1 & 1240& 1189/1193/1197  \\
$\Sigma^{*+}/\Sigma^{*0}/\Sigma^{*-}$& 3/2 & 1 & 1437 & 1383/1384/1387 \\
\hline
$\Xi^0/\Xi^-$ & 1/2 & 1/2 & 1350 & 1315/1322 \\
$\Xi^{*0}/\Xi^{*-}$ & 3/2 & 1/2 & 1559 & 1532/1535 \\
\hline
\end{tabular}
\end{table}

\begin{table}[t]
\caption{Masses (in MeV) with error bars for the sexaquark and $H$ dibaryon with the $SU_f(3)$ symmetry is broken. We kept the restriction $I^2$=0. Thresholds are calculated with the masses taken from Table \ref{barsin}.}
\label{sin}
\centering
\begin{tabular}{c c c c c}
\hline
$S$ & Sexaquark & $H$ dibaryon & Threshold ($I=0$) & Mass \\
\hline
0 & $2657\pm6$ & $2412\pm2$ & $\Lambda\Lambda$ & 2312 \\
1 & $2810\pm5$ & $2378\pm4$ & $\Lambda\Lambda$ & 2312 \\
2 & $2882\pm2$ & $2747\pm2$ & $N\Xi^*$ & 2566 \\
3 & -- & $3087\pm5$ & $\Sigma^*\Sigma^*$ & 2874 \\
\hline
\end{tabular}
\end{table}

The trends observed in Table \ref{sin} are similar to those discussed for Table \ref{masses}. As before, a $H$ dibaryon is systematically more stable than the sexaquark with the same quantum numbers, whenever the latter is allowed by symmetry. We also find that, although the thresholds are different, all configurations remain unbound with respect to the corresponding two-baryon limit with the same spin and isospin. 
The least unbound state is now the $H$ dibaryon with $S=1$, which lies approximately $70$ MeV above threshold, a larger excess than in the case with full $SU_f(3)$ symmetry. Therefore, flavor-symmetry breaking does not improve the binding. On the contrary, the closest configuration to threshold in Table \ref{sin} lies farther from it than the near-threshold states found in the fully symmetric calculation.

Overall, our results indicate that, within the present framework, compact six-quark configurations do not lead to bound states, and the system is consistently driven towards a two-baryon-like configuration.

\subsection{Structure}

From the results given above, we can confidently say that exploring the two possible Hilbert spaces for the $uuddss$ structure produce systems with distinctly different masses for the same set of quantum numbers.  In this section,
we analyze the connection between the masses and the spatial distributions of the quarks.  To do so, we took as a measurement of size the square root of the distances between pairs of quarks $ij$ squared, $\sqrt(r_ij^2)$.  For symmetry reasons, the distances between the first three quarks are always equal to each other within the error bars, and the same is true for the distances between the particles with the last three labels. On the other hand,  distances between quarks in the first "baryon" and the second one could be different in for a compact sexaquark and and $H$ dibaryon.  Those values are shown in Table \ref{distances} for both arrangements. 

\begin{table}[t]
\caption{Averaged distances (in fm), together with their error bars, for the sexaquark and $H$ dibaryon states for different $(S,F^2)$ sectors in the arrangements allowed by symmetry.  $\sqrt(r_{12}^2)$ stands for the averaged distances between all $ij$ pairs in the sexaquark (S) and the
distances between quarks in the same "baryon" in the $H$ dibaryon ($H$). $\sqrt(r_{14}^2)$ are the distances between pairs not inside the same "baryon". }
\label{distances}
\centering
\begin{tabular}{c c c c c}
\hline
$S$ & $F^2$ & $\sqrt(r_{12}^2)$ (S) & $\sqrt(r_{12}^2)$ ($H$) & $\sqrt(r_{14}^2)$ ($H$) \\
\hline
0 & 0 & $0.97\pm0.02$ &  $0.96\pm0.02$ & $0.95\pm0.02$ \\
0 & 3 & -- & $0.90\pm0.02$  & $2.56\pm0.02$ \\
0 & 8 & $1.08\pm0.02$ & $0.90\pm0.02$ & $2.62\pm0.04$  \\
\hline
1 & 0 & -- & $0.92\pm0.02$ & $1.15\pm0.03$ \\
1 & 3 & $1.03\pm0.02$ & $0.90\pm0.02$ &$0.94\pm0.02$  \\
1 & 8 & -- & $0.90\pm0.02$ & $2.34\pm0.03$ \\
\hline
2 & 0 & -- &$1.07\pm0.02$ & $1.07\pm0.02$  \\
2 & 3 & $1.05\pm0.02$ & $1.02\pm0.02$ & $1.02\pm0.02$ \\
2 & 8 & $1.12\pm0.02$ &$0.93\pm0.02$ & $2.91\pm0.04$ \\
\hline
3 & 0 & -- & $1.11\pm0.02$ & $1.11\pm0.02$ \\
3 & 3 & -- & $1.12\pm0.02$ &   $1.12\pm0.02$ \\
3 & 8 & -- & $1.07\pm0.02$ & $1.12\pm0.02$  \\
\hline
\end{tabular}
\end{table}

\begin{table}[h]
\caption{Averaged interquark distances (in fm) for the baryons considered when calculated the mass thresholds. The differences between pairs are always within the provided error bars}.  
\label{baryonsdis}
\centering
\begin{tabular}{c c c c c}
\hline
Baryon & $S$ & $F^2$ & $I$ &  $\sqrt(r_{12}^2)$ \\
\hline
$p/n$ & 1/2 & 3 & 1/2 &  $0.95\pm0.01$\\
$\Delta^+/ \Delta^0$ & 3/2 & 6 & 3/2 &  $1.09\pm0.02$ \\ 
\hline
$\Lambda^0$ & 1/2 & 3 & 0 & $0.90\pm0.01$  \\
$\Sigma^+/\Sigma^0/\Sigma^-$& 1/2 & 3 &1 &  $0.90\pm0.02$   \\
$\Sigma^{*+}/\Sigma^{*0}/\Sigma^{*-}$& 3/2 & 6 & 1 & $1.01\pm0.02$  \\
\hline
$\Xi^0/\Xi^-$ & 1/2 & 3 & 1/2 & $0.84\pm0.02$ \\
$\Xi^{*0}/\Xi^{*-}$ & 3/2 & 6 & 1/2 & $0.93\pm0.01$ \\
\hline
\end{tabular}
\end{table}

A close inspection of Table \ref{baryons} allows us to establish a rather remarkable thing: the baryon-baryon-like splitting of the $H$ dibaryon internal  wavefunction does not necessarily implies two units separated from one another.  For instance,  in the $S$=0,$F^2$=0 case, the distances between any of the first three quarks is similar to the distance between a particle in the first threesome and another at the second "baryon".  Moreover, that they are both equal, within the error bars, to the ones for the sexaquark with the same quantum numbers.  This is reasonable since both structures have also equal masses (see Table \ref{masses}).  Something similar can be said of all other cases with $F^2$=0.  

Interestingly, there are situations in which the two calculated distances between quarks in the $H$ dibaryon are clearly different for each other. Those are first $(S,F^2)=(0,3)$, $(0,8)$,  $(1,8)$, in which we have two baryons close in size to a $\Lambda$ baryon (see Table \ref{baryonsdis}), separated from each other a sizeable distance ($\sim$ 2.5 fm).  We have a similar situation for $(S,F^2)=(2,8)$,  that cannot be a $\Lambda \Lambda$ arrangement, and has a similar size as a $N\Xi^*$ combination.  
These results indicate that the internal part of the wavefunction does not uniquely determine the final structure of the multiquark system,  and that the DMC procedure is able to converge to the hexaquark with smaller mass compatible with a set of quantum numbers.  

When we compare the results in Tables \ref{masses} and \ref{distances}, we can see that hexaquark configurations that can be seen as two"baryons" loosely bounded are the ones with lower masses and closer to their respective baryon-baryon thresholds. 
However, one may think that we have in fact are two baryons artificially close to each other due to the character of the radial part of the trial wavefunction (Eq. \ref{eq:radialwf}).  This function includes correlations between all pairs of quarks, and could be though as incompatible with the splitting of a multiquark in two smaller units. This is not so, since the DMC technique is able to break the system into subunits when those subsystems are colorless.  This was shown for all heavy pentaquarks in Ref. \cite{Gordillo2024b}. 

 \section{Conclusions}

We have studied the $uuddss$ system within a constituent quark model using the AL1 interaction and a diffusion Monte Carlo algorithm.  Two types of configurations have been considered: fully antisymmetric compact six-quark states (sexaquarks) and baryon-baryon-like configurations associated with the $H$ dibaryon. The former internal function is an extension of the single configuration deduced in Ref. \cite{Farrar2023} but not used to obtain masses and structures.  The latter is also an evolution  from previous coupled channel calculations that include terms beyond combinations of direct $\Lambda \Lambda$, $N\Xi$ and $\Sigma \Sigma$ products \cite{Kodama1994,Haidenbauer2011,Carames2012}, in particular, hidden color. We tackled other spins apart from $S$=0 and considered both isospin and flavor degrees of freedom.  
The sexaquark and $H$-dibaryon configurations are treated on the same
footing, allowing a direct comparison of their masses within a common
framework. 

Within the standard nonrelativistic AL1 model used here, the
masses obtained for the sexaquark configurations are always above, or at
most comparable to, those of the $H$-dibaryon configurations with the same quantum numbers.
This comparison is meaningful because the two configurations arise from different antisymmetrization prescriptions, and therefore correspond to different wave functions. Even in the case that these wave functions were not orthogonal and could not be 
distinguished with certainty in a single measurement,  non-identical non-orthogonal states could still be probed statistically,  in the sense that
suitable measurements, such as the application of the adequate set of projectors,  may produce different probabilities or expectation
values~\cite{TongTopicsQM}. 
This does not make the two descriptions equivalent:
their different internal structures may lead to different masses,
interquark distances, correlation functions, or overlaps with
baryon--baryon channels. In this sense, the compact six-quark and
baryon--baryon-like character of the states can be assessed through
several observables.

On the other hand, to establish the absolute stability of both the
sexaquark and $H$-dibaryon configurations, their masses must be compared to the corresponding two-baryon thresholds. This comparison has been made in Table~\ref{masses}, using the baryon masses reported in Table~\ref{baryons}. Those baryon masses were calculated
without including any three-body correction, even though the original
description of the AL1 interaction in Ref.~\cite{SilvestreBrac1996}
introduced an ad hoc three-body term, with no direct physical
justification, in order to improve the fit to the experimental baryon masses. This correction is sizable, negative, and ranges from about $65$ MeV for the $p/n$ to about $20$ MeV for the
$\Xi$ baryons. Therefore, including such a term could modify the balance
between the $uuddss$ masses and their corresponding thresholds reported in Table~\ref{masses}.  The reason we did not consider those three-body terms in our calculations is that, even tough the are necessary to account correctly for the baryon masses, there is no standard recipe to apply them to larger multiquarks.  In fact, in the standard literature that term is typically ignored (see for instance Ref. \cite{vijande}). 

However, this is not the only way to consider three-body terms in the Hamiltonian, since one can introduce terms proportional to $\vec{\lambda}_{i}\cdot
\vec{\lambda}_{j}\cdot\vec{\lambda}_{k}$.  Few works have addressed genuine quark three-body interactions in six-quark 
systems using constituent quark model calculations \cite{stancu2002,park2015, park2024}.  In the first two cases, the three-body contributions to the hexaquark masses were found to be the same as for two separate baryons, 
which implies that, in those schematic models, such terms shift the
absolute six-quark mass and the two-baryon threshold by the same amount,
and therefore do not change the relative stability of hexaquarks with respect to the two-baryon thresholds.  On the other hand, 
recent exploratory calculations of quark three-body potentials in compact
dibaryons \cite{park2024} also indicate repulsive contributions in the dominant flavor
channels. Such contributions would not favor additional binding of the
compact $uuddss$ configuration, and are therefore consistent with the
absence of a bound compact state found here.
The same lines of reasoning could be applied to the flavor-symmetry breaking description of the hexaquarks, indicating that the absence of binding is a robust feature of the AL1 model.  

The analysis of interparticle distances provides a consistent picture of the internal structure of the system. Configurations that remain compact have larger masses, while those closer to the corresponding two-baryon thresholds show a clear tendency towards a baryon--baryon-like spatial structure.
The combination of the data on masses and distances obtained strongly indicate that,  within this approach, the $uuddss$ system as described by a non-relativistic AL1 potential,  does not favor compact configurations and tends instead towards two-baryon-like arrangements.

\begin{acknowledgments}
We acknowledge financial support from Ministerio Espa\~nol de Ciencia e Innovaci\'on under grant No.  PID2023-147469NB-C21 and Universidad Pablo de Olavide 
group GrIN-UPO FQM-205.
The author acknowledges, too, the use of the computer facilities of C3UPO at the Universidad Pablo de Olavide, de Sevilla within 
"Plan Propio de Investigación y Transferencia (2023-2026) de la UPO, por la Consejería de Universidades, Investigación e Innovación de la Junta de Andalucía y por el Programa FEDER Andalucía 2021-2027, 2023/00002/014
\end{acknowledgments}


\section{Apendix I}
The five functions corresponding to $F^2$=0 are:
\begin{equation}
\begin{aligned}
\zeta_f^{(1)} \;=\;&
-\frac{1}{\sqrt{6}}\,\chi_{f,1}^{(12)}
+\frac{1}{\sqrt{2}}\,\chi_{f,2}^{(12)}
-\frac{1}{\sqrt{6}}\,\chi_{f,1}^{(34)}
-\frac{1}{\sqrt{2}}\,\chi_{f,2}^{(34)}\\
&+\frac{1}{\sqrt{6}}\,\chi_{f,1}^{(13)}
+\frac{1}{\sqrt{2}}\,\chi_{f,2}^{(13)}
+\frac{1}{\sqrt{6}}\,\chi_{f,1}^{(14)}
+\frac{1}{\sqrt{2}}\,\chi_{f,2}^{(14)}\\
&-\frac{1}{\sqrt{6}}\,\chi_{f,1}^{(23)}
-\frac{1}{\sqrt{2}}\,\chi_{f,2}^{(23)}
+\frac{1}{\sqrt{6}}\,\chi_{f,1}^{(15)}
-\frac{1}{\sqrt{2}}\,\chi_{f,2}^{(15)}\\
&-\frac{1}{\sqrt{6}}\,\chi_{f,1}^{(16)}
+\frac{1}{\sqrt{2}}\,\chi_{f,2}^{(16)}
+\frac{1}{\sqrt{6}}\,\chi_{f,1}^{(45)}
+\frac{1}{\sqrt{2}}\,\chi_{f,2}^{(45)}\\
&-\frac{1}{\sqrt{6}}\,\chi_{f,1}^{(46)}
-\frac{1}{\sqrt{2}}\,\chi_{f,2}^{(46)} \,.
\end{aligned}
\end{equation}
\begin{equation}
\begin{aligned}
\zeta_f^{(2)} \;=\;&
-\frac{1}{\sqrt{6}}\,\chi_{f,1}^{(56)}
+\frac{1}{\sqrt{2}}\,\chi_{f,2}^{(56)}
+\frac{1}{\sqrt{6}}\,\chi_{f,1}^{(34)}
-\frac{1}{\sqrt{2}}\,\chi_{f,2}^{(34)}\\
&-\frac{1}{\sqrt{6}}\,\chi_{f,1}^{(14)}
+\frac{1}{\sqrt{2}}\,\chi_{f,2}^{(14)}
-\frac{1}{\sqrt{6}}\,\chi_{f,1}^{(15)}
-\frac{1}{\sqrt{2}}\,\chi_{f,2}^{(15)}\\
&+\frac{1}{\sqrt{6}}\,\chi_{f,1}^{(16)}
+\frac{1}{\sqrt{2}}\,\chi_{f,2}^{(16)}
-\frac{1}{\sqrt{6}}\,\chi_{f,1}^{(25)}
-\frac{1}{\sqrt{2}}\,\chi_{f,2}^{(25)}\\
&+\frac{1}{\sqrt{6}}\,\chi_{f,1}^{(35)}
+\frac{1}{\sqrt{2}}\,\chi_{f,2}^{(35)}
-\frac{1}{\sqrt{6}}\,\chi_{f,1}^{(36)}
-\frac{1}{\sqrt{2}}\,\chi_{f,2}^{(36)}\\
&+\frac{1}{\sqrt{6}}\,\chi_{f,1}^{(45)}
+\frac{1}{\sqrt{2}}\,\chi_{f,2}^{(45)} \,.
\end{aligned}
\end{equation}

\begin{equation}
\begin{aligned}
\zeta_f^{(3)} \;=\;&
-\sqrt{\frac{2}{3}}\,\chi_{f,1}^{(12)}
-\sqrt{\frac{2}{3}}\,\chi_{f,1}^{(56)}
+\frac{1}{\sqrt{6}}\,\chi_{f,1}^{(34)}
-\frac{1}{\sqrt{2}}\,\chi_{f,2}^{(34)}\\
&+\frac{1}{\sqrt{6}}\,\chi_{f,1}^{(13)}
+\frac{1}{\sqrt{2}}\,\chi_{f,2}^{(13)}
+\frac{1}{\sqrt{6}}\,\chi_{f,1}^{(14)}
+\frac{1}{\sqrt{2}}\,\chi_{f,2}^{(14)}\\
&-\frac{1}{\sqrt{6}}\,\chi_{f,1}^{(24)}
-\frac{1}{\sqrt{2}}\,\chi_{f,2}^{(24)}
+\frac{1}{\sqrt{6}}\,\chi_{f,1}^{(15)}
-\frac{1}{\sqrt{2}}\,\chi_{f,2}^{(15)}\\
&+\frac{1}{\sqrt{6}}\,\chi_{f,1}^{(16)}
+\frac{1}{\sqrt{2}}\,\chi_{f,2}^{(16)}
-\sqrt{\frac{2}{3}}\,\chi_{f,1}^{(25)}
+\frac{1}{\sqrt{6}}\,\chi_{f,1}^{(35)}\\
&+\frac{1}{\sqrt{2}}\,\chi_{f,2}^{(35)}
+\frac{1}{\sqrt{6}}\,\chi_{f,1}^{(45)}
+\frac{1}{\sqrt{2}}\,\chi_{f,2}^{(45)}
-\sqrt{\frac{2}{3}}\,\chi_{f,1}^{(46)} \,.
\end{aligned}
\end{equation}

\begin{equation}
\begin{aligned}
\zeta_f^{(4)} \;=\;&
-\frac{1}{\sqrt{6}}\,\chi_{f,1}^{(12)}
+\frac{1}{\sqrt{2}}\,\chi_{f,2}^{(12)}
-\frac{1}{\sqrt{6}}\,\chi_{f,1}^{(56)}
+\frac{1}{\sqrt{2}}\,\chi_{f,2}^{(56)}\\
&-\frac{1}{\sqrt{6}}\,\chi_{f,1}^{(34)}
-\frac{1}{\sqrt{2}}\,\chi_{f,2}^{(34)}
+\sqrt{\frac{2}{3}}\,\chi_{f,1}^{(13)}
+\frac{1}{\sqrt{6}}\,\chi_{f,1}^{(14)}\\
&+\frac{1}{\sqrt{2}}\,\chi_{f,2}^{(14)}
-\sqrt{\frac{2}{3}}\,\chi_{f,1}^{(23)}
-\frac{1}{\sqrt{6}}\,\chi_{f,1}^{(15)}
-\frac{1}{\sqrt{2}}\,\chi_{f,2}^{(15)} \\
&+\frac{1}{\sqrt{6}}\,\chi_{f,1}^{(16)}
+\frac{1}{\sqrt{2}}\,\chi_{f,2}^{(16)}
-\frac{1}{\sqrt{6}}\,\chi_{f,1}^{(26)}
-\frac{1}{\sqrt{2}}\,\chi_{f,2}^{(26)} \\
&+\sqrt{\frac{2}{3}}\,\chi_{f,1}^{(35)}
-\sqrt{\frac{2}{3}}\,\chi_{f,1}^{(36)}
+\frac{1}{\sqrt{6}}\,\chi_{f,1}^{(45)}
+\frac{1}{\sqrt{2}}\,\chi_{f,2}^{(45)} \,.
\end{aligned}
\end{equation}

\begin{equation}
\begin{aligned}
\zeta_f^{(5)} \;=\;&
-\sqrt{\frac{2}{3}}\,\chi_{f,1}^{(12)}
-\sqrt{\frac{2}{3}}\,\chi_{f,1}^{(56)}
+\sqrt{\frac{2}{3}}\,\chi_{f,1}^{(13)}
+\frac{1}{\sqrt{6}}\,\chi_{f,1}^{(14)} \\
&+\frac{1}{\sqrt{2}}\,\chi_{f,2}^{(14)}
-\sqrt{\frac{2}{3}}\,\chi_{f,1}^{(24)}
+\frac{1}{\sqrt{6}}\,\chi_{f,1}^{(16)}
+\frac{1}{\sqrt{2}}\,\chi_{f,2}^{(16)} \\
&-\sqrt{\frac{2}{3}}\,\chi_{f,1}^{(26)}
+\sqrt{\frac{2}{3}}\,\chi_{f,1}^{(35)}
-\frac{1}{\sqrt{6}}\,\chi_{f,1}^{(45)}
+\frac{1}{\sqrt{2}}\,\chi_{f,2}^{(45)} \,.
\end{aligned}
\end{equation}

The 16 functions corresponding to $F^2$=3 are:
\begin{equation}
\begin{aligned}
\upsilon_f^{(1)}\;=\;& \frac{\sqrt{10}}{10}\,\chi_{f,1}^{(12)} - \frac{\sqrt{30}}{30}\,\chi_{f,2}^{(12)} + \frac{\sqrt{10}}{10}\,\chi_{f,1}^{(13)} - \frac{\sqrt{30}}{30}\,\chi_{f,2}^{(13)} \\
& - \frac{\sqrt{10}}{20}\,\chi_{f,1}^{(14)} - \frac{\sqrt{30}}{60}\,\chi_{f,2}^{(14)} - \frac{\sqrt{10}}{20}\,\chi_{f,1}^{(15)} - \frac{\sqrt{30}}{60}\,\chi_{f,2}^{(15)} \\
& + \frac{\sqrt{10}}{10}\,\chi_{f,1}^{(16)} + \frac{\sqrt{30}}{30}\,\chi_{f,2}^{(16)} + \frac{\sqrt{10}}{10}\,\chi_{f,1}^{(23)} - \frac{\sqrt{30}}{30}\,\chi_{f,2}^{(23)} \\
& - \frac{\sqrt{10}}{20}\,\chi_{f,1}^{(24)} - \frac{\sqrt{30}}{60}\,\chi_{f,2}^{(24)} - \frac{\sqrt{10}}{20}\,\chi_{f,1}^{(25)} - \frac{\sqrt{30}}{60}\,\chi_{f,2}^{(25)} \\
& + \frac{\sqrt{10}}{10}\,\chi_{f,1}^{(26)} + \frac{\sqrt{30}}{30}\,\chi_{f,2}^{(26)} - \frac{\sqrt{10}}{20}\,\chi_{f,1}^{(34)} - \frac{\sqrt{30}}{60}\,\chi_{f,2}^{(34)} \\
& - \frac{\sqrt{10}}{20}\,\chi_{f,1}^{(35)} - \frac{\sqrt{30}}{60}\,\chi_{f,2}^{(35)} + \frac{\sqrt{10}}{10}\,\chi_{f,1}^{(36)} + \frac{\sqrt{30}}{30}\,\chi_{f,2}^{(36)}\,.
\end{aligned}
\end{equation}

\begin{equation}
\begin{aligned}
\upsilon_f^{(2)}\;=\;& \frac{\sqrt{5}}{20}\,\chi_{f,1}^{(12)} - \frac{\sqrt{15}}{12}\,\chi_{f,2}^{(12)} + \frac{\sqrt{5}}{40}\,\chi_{f,1}^{(13)} + \frac{\sqrt{15}}{24}\,\chi_{f,2}^{(13)} \\
& - \frac{\sqrt{5}}{8}\,\chi_{f,1}^{(14)} + \frac{7 \sqrt{15}}{120}\,\chi_{f,2}^{(14)} + \frac{\sqrt{5}}{40}\,\chi_{f,1}^{(15)} - \frac{\sqrt{15}}{24}\,\chi_{f,2}^{(15)} \\
& + \frac{\sqrt{5}}{20}\,\chi_{f,1}^{(16)} + \frac{\sqrt{15}}{12}\,\chi_{f,2}^{(16)} + \frac{\sqrt{5}}{40}\,\chi_{f,1}^{(23)} + \frac{\sqrt{15}}{24}\,\chi_{f,2}^{(23)} \\
& - \frac{\sqrt{5}}{8}\,\chi_{f,1}^{(24)} + \frac{7 \sqrt{15}}{120}\,\chi_{f,2}^{(24)} + \frac{\sqrt{5}}{40}\,\chi_{f,1}^{(25)} - \frac{\sqrt{15}}{24}\,\chi_{f,2}^{(25)} \\
& - \frac{\sqrt{5}}{20}\,\chi_{f,1}^{(26)} + \frac{\sqrt{15}}{12}\,\chi_{f,2}^{(26)} + \frac{\sqrt{5}}{10}\,\chi_{f,1}^{(34)} + \frac{\sqrt{15}}{30}\,\chi_{f,2}^{(34)} \\
& + \frac{\sqrt{5}}{40}\,\chi_{f,1}^{(35)} + \frac{\sqrt{15}}{120}\,\chi_{f,2}^{(35)} - \frac{\sqrt{5}}{20}\,\chi_{f,1}^{(36)} - \frac{\sqrt{15}}{60}\,\chi_{f,2}^{(36)} \\
& + \frac{3 \sqrt{5}}{40}\,\chi_{f,1}^{(45)} + \frac{\sqrt{15}}{40}\,\chi_{f,2}^{(45)} - \frac{3 \sqrt{5}}{20}\,\chi_{f,1}^{(46)} - \frac{\sqrt{15}}{20}\,\chi_{f,2}^{(46)}\,.
\end{aligned}
\end{equation}

\begin{equation}
\begin{aligned}
\upsilon_f^{(3)}\;=\;& \frac{\sqrt{15}}{40}\,\chi_{f,1}^{(13)} + \frac{\sqrt{5}}{8}\,\chi_{f,2}^{(13)} + \frac{\sqrt{15}}{40}\,\chi_{f,1}^{(14)} + \frac{\sqrt{5}}{8}\,\chi_{f,2}^{(14)} \\
& + \frac{\sqrt{15}}{40}\,\chi_{f,1}^{(15)} - \frac{\sqrt{5}}{40}\,\chi_{f,2}^{(15)} - \frac{\sqrt{15}}{20}\,\chi_{f,1}^{(16)} + \frac{\sqrt{5}}{20}\,\chi_{f,2}^{(16)} \\
& - \frac{\sqrt{15}}{40}\,\chi_{f,1}^{(23)} - \frac{\sqrt{5}}{8}\,\chi_{f,2}^{(23)} - \frac{\sqrt{15}}{40}\,\chi_{f,1}^{(24)} - \frac{\sqrt{5}}{8}\,\chi_{f,2}^{(24)} \\
& - \frac{\sqrt{15}}{40}\,\chi_{f,1}^{(25)} + \frac{\sqrt{5}}{40}\,\chi_{f,2}^{(25)} + \frac{\sqrt{15}}{20}\,\chi_{f,1}^{(26)} + \frac{\sqrt{5}}{20}\,\chi_{f,2}^{(26)} \\
& + \frac{\sqrt{15}}{20}\,\chi_{f,1}^{(34)} - \frac{3 \sqrt{5}}{20}\,\chi_{f,2}^{(34)} - \frac{\sqrt{15}}{40}\,\chi_{f,1}^{(35)} + \frac{3 \sqrt{5}}{40}\,\chi_{f,2}^{(35)} \\
& + \frac{\sqrt{15}}{20}\,\chi_{f,1}^{(36)} - \frac{3 \sqrt{5}}{20}\,\chi_{f,2}^{(36)} - \frac{\sqrt{15}}{40}\,\chi_{f,1}^{(45)} + \frac{3 \sqrt{5}}{40}\,\chi_{f,2}^{(45)} \\
& + \frac{\sqrt{15}}{20}\,\chi_{f,1}^{(46)} - \frac{3 \sqrt{5}}{20}\,\chi_{f,2}^{(46)}\,.
\end{aligned}
\end{equation}

\begin{equation}
\begin{aligned}
\upsilon_f^{(4)}\;=\;& \frac{\sqrt{15}}{20}\,\chi_{f,1}^{(12)} + \frac{\sqrt{5}}{20}\,\chi_{f,2}^{(12)} - \frac{\sqrt{15}}{40}\,\chi_{f,1}^{(13)} - \frac{\sqrt{5}}{40}\,\chi_{f,2}^{(13)} \\
& - \frac{\sqrt{15}}{40}\,\chi_{f,1}^{(14)} - \frac{\sqrt{5}}{40}\,\chi_{f,2}^{(14)} - \frac{3 \sqrt{15}}{40}\,\chi_{f,1}^{(15)} + \frac{3 \sqrt{5}}{40}\,\chi_{f,2}^{(15)} \\
& + \frac{\sqrt{15}}{20}\,\chi_{f,1}^{(16)} - \frac{\sqrt{5}}{20}\,\chi_{f,2}^{(16)} - \frac{\sqrt{15}}{40}\,\chi_{f,1}^{(23)} - \frac{\sqrt{5}}{40}\,\chi_{f,2}^{(23)} \\
& - \frac{\sqrt{15}}{40}\,\chi_{f,1}^{(24)} - \frac{\sqrt{5}}{40}\,\chi_{f,2}^{(24)} - \frac{3 \sqrt{15}}{40}\,\chi_{f,1}^{(25)} + \frac{3 \sqrt{5}}{40}\,\chi_{f,2}^{(25)} \\
& + \frac{\sqrt{15}}{20}\,\chi_{f,1}^{(26)} + \frac{\sqrt{5}}{20}\,\chi_{f,2}^{(26)} + \frac{\sqrt{15}}{20}\,\chi_{f,1}^{(34)} + \frac{\sqrt{5}}{20}\,\chi_{f,2}^{(34)} \\
& + \frac{3 \sqrt{15}}{40}\,\chi_{f,1}^{(35)} + \frac{3 \sqrt{5}}{40}\,\chi_{f,2}^{(35)} - \frac{\sqrt{15}}{20}\,\chi_{f,1}^{(36)} + \frac{\sqrt{5}}{20}\,\chi_{f,2}^{(36)} \\
& + \frac{3 \sqrt{15}}{40}\,\chi_{f,1}^{(45)} + \frac{3 \sqrt{5}}{40}\,\chi_{f,2}^{(45)} - \frac{\sqrt{15}}{20}\,\chi_{f,1}^{(46)} - \frac{\sqrt{5}}{20}\,\chi_{f,2}^{(46)} \\
& - \frac{\sqrt{15}}{10}\,\chi_{f,1}^{(56)} - \frac{\sqrt{5}}{10}\,\chi_{f,2}^{(56)}\,.
\end{aligned}
\end{equation}

\begin{equation}
\begin{aligned}
\upsilon_f^{(5)}\;=\;& \frac{3 \sqrt{5}}{40}\,\chi_{f,1}^{(13)} + \frac{\sqrt{15}}{40}\,\chi_{f,2}^{(13)} - \frac{3 \sqrt{5}}{40}\,\chi_{f,1}^{(14)} - \frac{\sqrt{15}}{40}\,\chi_{f,2}^{(14)} \\
& - \frac{3 \sqrt{5}}{40}\,\chi_{f,1}^{(15)} - \frac{3 \sqrt{15}}{40}\,\chi_{f,2}^{(15)} + \frac{\sqrt{5}}{20}\,\chi_{f,1}^{(16)} + \frac{\sqrt{15}}{20}\,\chi_{f,2}^{(16)} \\
& - \frac{3 \sqrt{5}}{40}\,\chi_{f,1}^{(23)} - \frac{\sqrt{15}}{40}\,\chi_{f,2}^{(23)} + \frac{3 \sqrt{5}}{40}\,\chi_{f,1}^{(24)} + \frac{\sqrt{15}}{40}\,\chi_{f,2}^{(24)} \\
& + \frac{3 \sqrt{5}}{40}\,\chi_{f,1}^{(25)} + \frac{3 \sqrt{15}}{40}\,\chi_{f,2}^{(25)} + \frac{\sqrt{5}}{20}\,\chi_{f,1}^{(26)} - \frac{\sqrt{15}}{20}\,\chi_{f,2}^{(26)} \\
& - \frac{3 \sqrt{5}}{40}\,\chi_{f,1}^{(35)} + \frac{3 \sqrt{15}}{40}\,\chi_{f,2}^{(35)} + \frac{\sqrt{5}}{20}\,\chi_{f,1}^{(36)} - \frac{\sqrt{15}}{20}\,\chi_{f,2}^{(36)} \\
& + \frac{3 \sqrt{5}}{40}\,\chi_{f,1}^{(45)} - \frac{3 \sqrt{15}}{40}\,\chi_{f,2}^{(45)} - \frac{\sqrt{5}}{20}\,\chi_{f,1}^{(46)} + \frac{\sqrt{15}}{20}\,\chi_{f,2}^{(46)} \\
& - \frac{\sqrt{5}}{10}\,\chi_{f,1}^{(56)} + \frac{\sqrt{15}}{10}\,\chi_{f,2}^{(56)}\,.
\end{aligned}
\end{equation}

\begin{equation}
\begin{aligned}
\upsilon_f^{(6)}\;=\;& - \frac{\sqrt{30}}{30}\,\chi_{f,1}^{(12)} - \frac{\sqrt{10}}{10}\,\chi_{f,2}^{(12)} - \frac{\sqrt{30}}{30}\,\chi_{f,1}^{(13)} - \frac{\sqrt{10}}{10}\,\chi_{f,2}^{(13)} \\
& + \frac{\sqrt{30}}{20}\,\chi_{f,1}^{(14)} + \frac{\sqrt{10}}{20}\,\chi_{f,2}^{(14)} - \frac{\sqrt{30}}{20}\,\chi_{f,1}^{(15)} - \frac{\sqrt{10}}{20}\,\chi_{f,2}^{(15)} \\
& - \frac{\sqrt{30}}{30}\,\chi_{f,1}^{(23)} - \frac{\sqrt{10}}{10}\,\chi_{f,2}^{(23)} + \frac{\sqrt{30}}{20}\,\chi_{f,1}^{(24)} + \frac{\sqrt{10}}{20}\,\chi_{f,2}^{(24)} \\
& - \frac{\sqrt{30}}{20}\,\chi_{f,1}^{(25)} - \frac{\sqrt{10}}{20}\,\chi_{f,2}^{(25)} + \frac{\sqrt{30}}{20}\,\chi_{f,1}^{(34)} + \frac{\sqrt{10}}{20}\,\chi_{f,2}^{(34)} \\
& - \frac{\sqrt{30}}{20}\,\chi_{f,1}^{(35)} - \frac{\sqrt{10}}{20}\,\chi_{f,2}^{(35)}\,.
\end{aligned}
\end{equation}

\begin{equation}
\begin{aligned}
\upsilon_f^{(7)}\;=\;& \frac{7 \sqrt{15}}{60}\,\chi_{f,1}^{(12)} + \frac{\sqrt{5}}{20}\,\chi_{f,2}^{(12)} - \frac{7 \sqrt{15}}{120}\,\chi_{f,1}^{(13)} - \frac{\sqrt{5}}{40}\,\chi_{f,2}^{(13)} \\
& + \frac{\sqrt{15}}{40}\,\chi_{f,1}^{(14)} + \frac{\sqrt{5}}{8}\,\chi_{f,2}^{(14)} + \frac{\sqrt{15}}{40}\,\chi_{f,1}^{(15)} - \frac{\sqrt{5}}{8}\,\chi_{f,2}^{(15)} \\
& - \frac{7 \sqrt{15}}{120}\,\chi_{f,1}^{(23)} - \frac{\sqrt{5}}{40}\,\chi_{f,2}^{(23)} + \frac{\sqrt{15}}{40}\,\chi_{f,1}^{(24)} + \frac{\sqrt{5}}{8}\,\chi_{f,2}^{(24)} \\
& + \frac{\sqrt{15}}{40}\,\chi_{f,1}^{(25)} - \frac{\sqrt{5}}{8}\,\chi_{f,2}^{(25)} - \frac{\sqrt{15}}{10}\,\chi_{f,1}^{(34)} - \frac{\sqrt{5}}{10}\,\chi_{f,2}^{(34)} \\
& + \frac{\sqrt{15}}{40}\,\chi_{f,1}^{(35)} + \frac{\sqrt{5}}{40}\,\chi_{f,2}^{(35)} + \frac{3 \sqrt{15}}{40}\,\chi_{f,1}^{(45)} + \frac{3 \sqrt{5}}{40}\,\chi_{f,2}^{(45)}\,.
\end{aligned}
\end{equation}

\begin{equation}
\begin{aligned}
\upsilon_f^{(8)}\;=\;& - \frac{7 \sqrt{5}}{40}\,\chi_{f,1}^{(13)} - \frac{\sqrt{15}}{40}\,\chi_{f,2}^{(13)} - \frac{7 \sqrt{5}}{40}\,\chi_{f,1}^{(14)} - \frac{\sqrt{15}}{40}\,\chi_{f,2}^{(14)} \\
& + \frac{3 \sqrt{5}}{40}\,\chi_{f,1}^{(15)} - \frac{\sqrt{15}}{40}\,\chi_{f,2}^{(15)} + \frac{7 \sqrt{5}}{40}\,\chi_{f,1}^{(23)} + \frac{\sqrt{15}}{40}\,\chi_{f,2}^{(23)} \\
& + \frac{7 \sqrt{5}}{40}\,\chi_{f,1}^{(24)} + \frac{\sqrt{15}}{40}\,\chi_{f,2}^{(24)} - \frac{3 \sqrt{5}}{40}\,\chi_{f,1}^{(25)} + \frac{\sqrt{15}}{40}\,\chi_{f,2}^{(25)} \\
& + \frac{\sqrt{5}}{20}\,\chi_{f,1}^{(34)} - \frac{\sqrt{15}}{20}\,\chi_{f,2}^{(34)} - \frac{3 \sqrt{5}}{40}\,\chi_{f,1}^{(35)} + \frac{3 \sqrt{15}}{40}\,\chi_{f,2}^{(35)} \\
& - \frac{3 \sqrt{5}}{40}\,\chi_{f,1}^{(45)} + \frac{3 \sqrt{15}}{40}\,\chi_{f,2}^{(45)}\,.
\end{aligned}
\end{equation}

\begin{equation}
\begin{aligned}
\upsilon_f^{(9)}\;=\;& - \frac{1}{4}\,\chi_{f,1}^{(12)} + \frac{\sqrt{3}}{4}\,\chi_{f,2}^{(12)} + \frac{1}{8}\,\chi_{f,1}^{(13)} - \frac{\sqrt{3}}{8}\,\chi_{f,2}^{(13)} \\
& - \frac{1}{8}\,\chi_{f,1}^{(14)} + \frac{\sqrt{3}}{8}\,\chi_{f,2}^{(14)} - \frac{1}{8}\,\chi_{f,1}^{(15)} - \frac{\sqrt{3}}{8}\,\chi_{f,2}^{(15)} \\
& + \frac{1}{8}\,\chi_{f,1}^{(23)} - \frac{\sqrt{3}}{8}\,\chi_{f,2}^{(23)} - \frac{1}{8}\,\chi_{f,1}^{(24)} + \frac{\sqrt{3}}{8}\,\chi_{f,2}^{(24)} \\
& - \frac{1}{8}\,\chi_{f,1}^{(25)} - \frac{\sqrt{3}}{8}\,\chi_{f,2}^{(25)} + \frac{3}{8}\,\chi_{f,1}^{(35)} + \frac{\sqrt{3}}{8}\,\chi_{f,2}^{(35)} \\
& - \frac{3}{8}\,\chi_{f,1}^{(45)} - \frac{\sqrt{3}}{8}\,\chi_{f,2}^{(45)}\,.
\end{aligned}
\end{equation}

\begin{equation}
\begin{aligned}
\upsilon_f^{(10)}\;=\;& - \frac{\sqrt{3}}{8}\,\chi_{f,1}^{(13)} + \frac{3}{8}\,\chi_{f,2}^{(13)} + \frac{\sqrt{3}}{24}\,\chi_{f,1}^{(14)} - \frac{1}{8}\,\chi_{f,2}^{(14)} \\
& - \frac{5 \sqrt{3}}{24}\,\chi_{f,1}^{(15)} - \frac{1}{8}\,\chi_{f,2}^{(15)} + \frac{\sqrt{3}}{8}\,\chi_{f,1}^{(23)} - \frac{3}{8}\,\chi_{f,2}^{(23)} \\
& - \frac{\sqrt{3}}{24}\,\chi_{f,1}^{(24)} + \frac{1}{8}\,\chi_{f,2}^{(24)} + \frac{5 \sqrt{3}}{24}\,\chi_{f,1}^{(25)} + \frac{1}{8}\,\chi_{f,2}^{(25)} \\
& - \frac{\sqrt{3}}{12}\,\chi_{f,1}^{(34)} + \frac{1}{4}\,\chi_{f,2}^{(34)} + \frac{\sqrt{3}}{24}\,\chi_{f,1}^{(35)} - \frac{1}{8}\,\chi_{f,2}^{(35)} \\
& - \frac{\sqrt{3}}{8}\,\chi_{f,1}^{(45)} + \frac{3}{8}\,\chi_{f,2}^{(45)}\,.
\end{aligned}
\end{equation}

\begin{equation}
\begin{aligned}
\upsilon_f^{(11)}\;=\;& - \frac{\sqrt{6}}{12}\,\chi_{f,1}^{(14)} + \frac{\sqrt{2}}{4}\,\chi_{f,2}^{(14)} - \frac{\sqrt{6}}{12}\,\chi_{f,1}^{(15)} + \frac{\sqrt{2}}{4}\,\chi_{f,2}^{(15)} \\
& + \frac{\sqrt{6}}{12}\,\chi_{f,1}^{(24)} - \frac{\sqrt{2}}{4}\,\chi_{f,2}^{(24)} + \frac{\sqrt{6}}{12}\,\chi_{f,1}^{(25)} - \frac{\sqrt{2}}{4}\,\chi_{f,2}^{(25)} \\
& - \frac{\sqrt{6}}{12}\,\chi_{f,1}^{(34)} + \frac{\sqrt{2}}{4}\,\chi_{f,2}^{(34)} - \frac{\sqrt{6}}{12}\,\chi_{f,1}^{(35)} + \frac{\sqrt{2}}{4}\,\chi_{f,2}^{(35)}\,.
\end{aligned}
\end{equation}

\begin{equation}
\begin{aligned}
\upsilon_f^{(12)}\;=\;& \frac{1}{4}\,\chi_{f,1}^{(12)} + \frac{\sqrt{3}}{12}\,\chi_{f,2}^{(12)} - \frac{1}{8}\,\chi_{f,1}^{(13)} - \frac{\sqrt{3}}{24}\,\chi_{f,2}^{(13)} \\
& - \frac{1}{8}\,\chi_{f,1}^{(14)} - \frac{\sqrt{3}}{24}\,\chi_{f,2}^{(14)} + \frac{1}{8}\,\chi_{f,1}^{(15)} - \frac{\sqrt{3}}{24}\,\chi_{f,2}^{(15)} \\
& - \frac{1}{4}\,\chi_{f,1}^{(16)} + \frac{\sqrt{3}}{12}\,\chi_{f,2}^{(16)} - \frac{1}{8}\,\chi_{f,1}^{(23)} - \frac{\sqrt{3}}{24}\,\chi_{f,2}^{(23)} \\
& - \frac{1}{8}\,\chi_{f,1}^{(24)} - \frac{\sqrt{3}}{24}\,\chi_{f,2}^{(24)} + \frac{1}{8}\,\chi_{f,1}^{(25)} - \frac{\sqrt{3}}{24}\,\chi_{f,2}^{(25)} \\
& - \frac{1}{4}\,\chi_{f,1}^{(26)} + \frac{\sqrt{3}}{12}\,\chi_{f,2}^{(26)} + \frac{1}{4}\,\chi_{f,1}^{(34)} + \frac{\sqrt{3}}{12}\,\chi_{f,2}^{(34)} \\
& - \frac{1}{8}\,\chi_{f,1}^{(35)} - \frac{\sqrt{3}}{24}\,\chi_{f,2}^{(35)} + \frac{1}{4}\,\chi_{f,1}^{(36)} + \frac{\sqrt{3}}{12}\,\chi_{f,2}^{(36)} \\
& - \frac{1}{8}\,\chi_{f,1}^{(45)} - \frac{\sqrt{3}}{24}\,\chi_{f,2}^{(45)} + \frac{1}{4}\,\chi_{f,1}^{(46)} + \frac{\sqrt{3}}{12}\,\chi_{f,2}^{(46)} \\
& - \frac{1}{2}\,\chi_{f,1}^{(56)} - \frac{\sqrt{3}}{6}\,\chi_{f,2}^{(56)}\,.
\end{aligned}
\end{equation}

\begin{equation}
\begin{aligned}
\upsilon_f^{(13)}\;=\;& - \frac{\sqrt{3}}{8}\,\chi_{f,1}^{(13)} - \frac{1}{8}\,\chi_{f,2}^{(13)} + \frac{\sqrt{3}}{8}\,\chi_{f,1}^{(14)} + \frac{1}{8}\,\chi_{f,2}^{(14)} \\
& - \frac{\sqrt{3}}{24}\,\chi_{f,1}^{(15)} - \frac{1}{8}\,\chi_{f,2}^{(15)} + \frac{\sqrt{3}}{12}\,\chi_{f,1}^{(16)} + \frac{1}{4}\,\chi_{f,2}^{(16)} \\
& + \frac{\sqrt{3}}{8}\,\chi_{f,1}^{(23)} + \frac{1}{8}\,\chi_{f,2}^{(23)} - \frac{\sqrt{3}}{8}\,\chi_{f,1}^{(24)} - \frac{1}{8}\,\chi_{f,2}^{(24)} \\
& + \frac{\sqrt{3}}{24}\,\chi_{f,1}^{(25)} + \frac{1}{8}\,\chi_{f,2}^{(25)} - \frac{\sqrt{3}}{12}\,\chi_{f,1}^{(26)} - \frac{1}{4}\,\chi_{f,2}^{(26)} \\
& - \frac{\sqrt{3}}{24}\,\chi_{f,1}^{(35)} + \frac{1}{8}\,\chi_{f,2}^{(35)} + \frac{\sqrt{3}}{12}\,\chi_{f,1}^{(36)} - \frac{1}{4}\,\chi_{f,2}^{(36)} \\
& + \frac{\sqrt{3}}{24}\,\chi_{f,1}^{(45)} - \frac{1}{8}\,\chi_{f,2}^{(45)} - \frac{\sqrt{3}}{12}\,\chi_{f,1}^{(46)} + \frac{1}{4}\,\chi_{f,2}^{(46)} \\
& + \frac{\sqrt{3}}{6}\,\chi_{f,1}^{(56)} - \frac{1}{2}\,\chi_{f,2}^{(56)}\,.
\end{aligned}
\end{equation}

\begin{equation}
\begin{aligned}
\upsilon_f^{(14)}\;=\;& - \frac{\sqrt{3}}{12}\,\chi_{f,1}^{(12)} + \frac{1}{4}\,\chi_{f,2}^{(12)} + \frac{\sqrt{3}}{24}\,\chi_{f,1}^{(13)} - \frac{1}{8}\,\chi_{f,2}^{(13)} \\
& - \frac{\sqrt{3}}{24}\,\chi_{f,1}^{(14)} + \frac{1}{8}\,\chi_{f,2}^{(14)} + \frac{\sqrt{3}}{24}\,\chi_{f,1}^{(15)} + \frac{1}{8}\,\chi_{f,2}^{(15)} \\
& + \frac{\sqrt{3}}{12}\,\chi_{f,1}^{(16)} + \frac{1}{4}\,\chi_{f,2}^{(16)} + \frac{\sqrt{3}}{24}\,\chi_{f,1}^{(23)} - \frac{1}{8}\,\chi_{f,2}^{(23)} \\
& - \frac{\sqrt{3}}{24}\,\chi_{f,1}^{(24)} + \frac{1}{8}\,\chi_{f,2}^{(24)} + \frac{\sqrt{3}}{24}\,\chi_{f,1}^{(25)} + \frac{1}{8}\,\chi_{f,2}^{(25)} \\
& + \frac{\sqrt{3}}{12}\,\chi_{f,1}^{(26)} + \frac{1}{4}\,\chi_{f,2}^{(26)} - \frac{\sqrt{3}}{8}\,\chi_{f,1}^{(35)} - \frac{1}{8}\,\chi_{f,2}^{(35)} \\
& - \frac{\sqrt{3}}{4}\,\chi_{f,1}^{(36)} - \frac{1}{4}\,\chi_{f,2}^{(36)} + \frac{\sqrt{3}}{8}\,\chi_{f,1}^{(45)} + \frac{1}{8}\,\chi_{f,2}^{(45)} \\
& + \frac{\sqrt{3}}{4}\,\chi_{f,1}^{(46)} + \frac{1}{4}\,\chi_{f,2}^{(46)}\,.
\end{aligned}
\end{equation}

\begin{equation}
\begin{aligned}
\upsilon_f^{(15)}\;=\;& \frac{1}{8}\,\chi_{f,1}^{(13)} - \frac{\sqrt{3}}{8}\,\chi_{f,2}^{(13)} - \frac{1}{24}\,\chi_{f,1}^{(14)} + \frac{\sqrt{3}}{24}\,\chi_{f,2}^{(14)} \\
& - \frac{5}{24}\,\chi_{f,1}^{(15)} - \frac{\sqrt{3}}{24}\,\chi_{f,2}^{(15)} - \frac{5}{12}\,\chi_{f,1}^{(16)} - \frac{\sqrt{3}}{12}\,\chi_{f,2}^{(16)} \\
& - \frac{1}{8}\,\chi_{f,1}^{(23)} + \frac{\sqrt{3}}{8}\,\chi_{f,2}^{(23)} + \frac{1}{24}\,\chi_{f,1}^{(24)} - \frac{\sqrt{3}}{24}\,\chi_{f,2}^{(24)} \\
& + \frac{5}{24}\,\chi_{f,1}^{(25)} + \frac{\sqrt{3}}{24}\,\chi_{f,2}^{(25)} + \frac{5}{12}\,\chi_{f,1}^{(26)} + \frac{\sqrt{3}}{12}\,\chi_{f,2}^{(26)} \\
& + \frac{1}{12}\,\chi_{f,1}^{(34)} - \frac{\sqrt{3}}{12}\,\chi_{f,2}^{(34)} + \frac{1}{24}\,\chi_{f,1}^{(35)} - \frac{\sqrt{3}}{24}\,\chi_{f,2}^{(35)} \\
& + \frac{1}{12}\,\chi_{f,1}^{(36)} - \frac{\sqrt{3}}{12}\,\chi_{f,2}^{(36)} - \frac{1}{8}\,\chi_{f,1}^{(45)} + \frac{\sqrt{3}}{8}\,\chi_{f,2}^{(45)} \\
& - \frac{1}{4}\,\chi_{f,1}^{(46)} + \frac{\sqrt{3}}{4}\,\chi_{f,2}^{(46)}\,.
\end{aligned}
\end{equation}

\begin{equation}
\begin{aligned}
\upsilon_f^{(16)}\;=\;& - \frac{\sqrt{2}}{12}\,\chi_{f,1}^{(14)} + \frac{\sqrt{6}}{12}\,\chi_{f,2}^{(14)} + \frac{\sqrt{2}}{12}\,\chi_{f,1}^{(15)} - \frac{\sqrt{6}}{12}\,\chi_{f,2}^{(15)} \\
& + \frac{\sqrt{2}}{6}\,\chi_{f,1}^{(16)} - \frac{\sqrt{6}}{6}\,\chi_{f,2}^{(16)} + \frac{\sqrt{2}}{12}\,\chi_{f,1}^{(24)} - \frac{\sqrt{6}}{12}\,\chi_{f,2}^{(24)} \\
& - \frac{\sqrt{2}}{12}\,\chi_{f,1}^{(25)} + \frac{\sqrt{6}}{12}\,\chi_{f,2}^{(25)} - \frac{\sqrt{2}}{6}\,\chi_{f,1}^{(26)} + \frac{\sqrt{6}}{6}\,\chi_{f,2}^{(26)} \\
& - \frac{\sqrt{2}}{12}\,\chi_{f,1}^{(34)} + \frac{\sqrt{6}}{12}\,\chi_{f,2}^{(34)} + \frac{\sqrt{2}}{12}\,\chi_{f,1}^{(35)} - \frac{\sqrt{6}}{12}\,\chi_{f,2}^{(35)} \\
& + \frac{\sqrt{2}}{6}\,\chi_{f,1}^{(36)} - \frac{\sqrt{6}}{6}\,\chi_{f,2}^{(36)}\,.
\end{aligned}
\end{equation}

And the nine functions for $F^2$=8 are:
\begin{equation}
\begin{aligned}
\tau_f^{(1)} \;=\;& - \sqrt{\frac{1}{8}}\,\chi_{f,1}^{(12)} - \sqrt{\frac{1}{24}}\,\chi_{f,2}^{(12)} - \sqrt{\frac{1}{8}}\,\chi_{f,1}^{(13)} - \sqrt{\frac{1}{24}}\,\chi_{f,2}^{(13)}\\
& - \sqrt{\frac{1}{8}}\,\chi_{f,1}^{(14)} - \sqrt{\frac{1}{24}}\,\chi_{f,2}^{(14)} - \sqrt{\frac{1}{8}}\,\chi_{f,1}^{(23)} - \sqrt{\frac{1}{24}}\,\chi_{f,2}^{(23)}\\
& - \sqrt{\frac{1}{8}}\,\chi_{f,1}^{(24)} - \sqrt{\frac{1}{24}}\,\chi_{f,2}^{(24)} - \sqrt{\frac{1}{8}}\,\chi_{f,1}^{(34)} - \sqrt{\frac{1}{24}}\,\chi_{f,2}^{(34)}
\end{aligned}
\end{equation}

\begin{equation}
\begin{aligned}
\tau_f^{(2)} \;=\;& 
+ \sqrt{\frac{1}{120}}\,\chi_{f,1}^{(12)}
- \sqrt{\frac{49}{360}}\,\chi_{f,2}^{(12)}
+ \sqrt{\frac{1}{120}}\,\chi_{f,1}^{(13)} \\
& - \sqrt{\frac{49}{360}}\,\chi_{f,2}^{(13)}
+ \sqrt{\frac{1}{120}}\,\chi_{f,1}^{(14)}
+ \sqrt{\frac{1}{360}}\,\chi_{f,2}^{(14)} \\
& + \sqrt{\frac{2}{15}}\,\chi_{f,1}^{(15)}
+ \sqrt{\frac{2}{45}}\,\chi_{f,2}^{(15)}
+ \sqrt{\frac{1}{120}}\,\chi_{f,1}^{(23)} \\
& - \sqrt{\frac{49}{360}}\,\chi_{f,2}^{(23)}
+ \sqrt{\frac{1}{120}}\,\chi_{f,1}^{(24)}
+ \sqrt{\frac{1}{360}}\,\chi_{f,2}^{(24)} \\
& + \sqrt{\frac{2}{15}}\,\chi_{f,1}^{(25)}
+ \sqrt{\frac{2}{45}}\,\chi_{f,2}^{(25)}
+ \sqrt{\frac{1}{120}}\,\chi_{f,1}^{(34)} \\
& + \sqrt{\frac{1}{360}}\,\chi_{f,2}^{(34)}
+ \sqrt{\frac{2}{15}}\,\chi_{f,1}^{(35)}
+ \sqrt{\frac{2}{45}}\,\chi_{f,2}^{(35)}
\end{aligned}
\end{equation}

\begin{equation}
\begin{aligned}
\tau_f^{(3)} \;=\;& 
- \sqrt{\frac{1}{15}}\,\chi_{f,1}^{(12)}
+ \sqrt{\frac{1}{45}}\,\chi_{f,2}^{(12)}
- \sqrt{\frac{1}{15}}\,\chi_{f,1}^{(13)} \\
& + \sqrt{\frac{1}{45}}\,\chi_{f,2}^{(13)}
+ \sqrt{\frac{1}{60}}\,\chi_{f,1}^{(14)}
+ \sqrt{\frac{1}{180}}\,\chi_{f,2}^{(14)} \\
& + \sqrt{\frac{1}{60}}\,\chi_{f,1}^{(15)}
+ \sqrt{\frac{1}{180}}\,\chi_{f,2}^{(15)}
+ \sqrt{\frac{3}{20}}\,\chi_{f,1}^{(16)} \\
& + \sqrt{\frac{1}{20}}\,\chi_{f,2}^{(16)}
- \sqrt{\frac{1}{15}}\,\chi_{f,1}^{(23)}
+ \sqrt{\frac{1}{45}}\,\chi_{f,2}^{(23)} \\
& + \sqrt{\frac{1}{60}}\,\chi_{f,1}^{(24)}
+ \sqrt{\frac{1}{180}}\,\chi_{f,2}^{(24)}
+ \sqrt{\frac{1}{60}}\,\chi_{f,1}^{(25)} \\
& + \sqrt{\frac{1}{180}}\,\chi_{f,2}^{(25)}
+ \sqrt{\frac{3}{20}}\,\chi_{f,1}^{(26)}
+ \sqrt{\frac{1}{20}}\,\chi_{f,2}^{(26)} \\
& + \sqrt{\frac{1}{60}}\,\chi_{f,1}^{(34)}
+ \sqrt{\frac{1}{180}}\,\chi_{f,2}^{(34)}
+ \sqrt{\frac{1}{60}}\,\chi_{f,1}^{(35)} \\
& + \sqrt{\frac{1}{180}}\,\chi_{f,2}^{(35)}
+ \sqrt{\frac{3}{20}}\,\chi_{f,1}^{(36)}
+ \sqrt{\frac{1}{20}}\,\chi_{f,2}^{(36)}
\end{aligned}
\end{equation}

\begin{equation}
\begin{aligned}
\tau_f^{(4)} \;=\;& 
- \sqrt{\frac{1}{15}}\,\chi_{f,1}^{(12)}
+ \sqrt{\frac{1}{45}}\,\chi_{f,2}^{(12)}
+ \sqrt{\frac{1}{60}}\,\chi_{f,1}^{(13)} \\
& - \sqrt{\frac{1}{180}}\,\chi_{f,2}^{(13)}
+ \sqrt{\frac{1}{60}}\,\chi_{f,1}^{(14)}
- \sqrt{\frac{5}{36}}\,\chi_{f,2}^{(14)} \\
& + \sqrt{\frac{1}{60}}\,\chi_{f,1}^{(15)}
- \sqrt{\frac{5}{36}}\,\chi_{f,2}^{(15)}
+ \sqrt{\frac{1}{60}}\,\chi_{f,1}^{(23)} \\
& - \sqrt{\frac{1}{180}}\,\chi_{f,2}^{(23)}
+ \sqrt{\frac{1}{60}}\,\chi_{f,1}^{(24)}
- \sqrt{\frac{5}{36}}\,\chi_{f,2}^{(24)} \\
& + \sqrt{\frac{1}{60}}\,\chi_{f,1}^{(25)}
- \sqrt{\frac{5}{36}}\,\chi_{f,2}^{(25)}
+ \sqrt{\frac{1}{60}}\,\chi_{f,1}^{(34)} \\
& + \sqrt{\frac{1}{180}}\,\chi_{f,2}^{(34)}
+ \sqrt{\frac{1}{60}}\,\chi_{f,1}^{(35)}
+ \sqrt{\frac{1}{180}}\,\chi_{f,2}^{(35)} \\
& + \sqrt{\frac{3}{20}}\,\chi_{f,1}^{(45)}
+ \sqrt{\frac{1}{20}}\,\chi_{f,2}^{(45)}
\end{aligned}
\end{equation}

\begin{equation}
\begin{aligned}
\tau_f^{(5)} \;=\;& 
- \sqrt{\frac{1}{20}}\,\chi_{f,1}^{(13)}
+ \sqrt{\frac{1}{60}}\,\chi_{f,2}^{(13)}
- \sqrt{\frac{1}{20}}\,\chi_{f,1}^{(14)} \\
& + \sqrt{\frac{1}{60}}\,\chi_{f,2}^{(14)}
- \sqrt{\frac{1}{20}}\,\chi_{f,1}^{(15)}
+ \sqrt{\frac{1}{60}}\,\chi_{f,2}^{(15)} \\
& + \sqrt{\frac{1}{20}}\,\chi_{f,1}^{(23)}
- \sqrt{\frac{1}{60}}\,\chi_{f,2}^{(23)}
+ \sqrt{\frac{1}{20}}\,\chi_{f,1}^{(24)} \\
& - \sqrt{\frac{1}{60}}\,\chi_{f,2}^{(24)}
+ \sqrt{\frac{1}{20}}\,\chi_{f,1}^{(25)}
- \sqrt{\frac{1}{60}}\,\chi_{f,2}^{(25)} \\
& + \sqrt{\frac{1}{20}}\,\chi_{f,1}^{(34)}
- \sqrt{\frac{3}{20}}\,\chi_{f,2}^{(34)}
+ \sqrt{\frac{1}{20}}\,\chi_{f,1}^{(35)} \\
& - \sqrt{\frac{3}{20}}\,\chi_{f,2}^{(35)}
+ \sqrt{\frac{1}{20}}\,\chi_{f,1}^{(45)}
- \sqrt{\frac{3}{20}}\,\chi_{f,2}^{(45)}
\end{aligned}
\end{equation}

\begin{equation}
\begin{aligned}
\tau_f^{(6)} \;=\;& 
- \sqrt{\frac{1}{120}}\,\chi_{f,1}^{(12)}
- \sqrt{\frac{5}{72}}\,\chi_{f,2}^{(12)}
+ \sqrt{\frac{1}{480}}\,\chi_{f,1}^{(13)} \\
& + \sqrt{\frac{5}{288}}\,\chi_{f,2}^{(13)}
- \sqrt{\frac{5}{96}}\,\chi_{f,1}^{(14)}
+ \sqrt{\frac{49}{1440}}\,\chi_{f,2}^{(14)} \\
& + \sqrt{\frac{1}{480}}\,\chi_{f,1}^{(15)}
- \sqrt{\frac{5}{288}}\,\chi_{f,2}^{(15)}
+ \sqrt{\frac{3}{160}}\,\chi_{f,1}^{(16)} \\
& - \sqrt{\frac{5}{32}}\,\chi_{f,2}^{(16)}
+ \sqrt{\frac{1}{480}}\,\chi_{f,1}^{(23)}
+ \sqrt{\frac{5}{288}}\,\chi_{f,2}^{(23)} \\
& - \sqrt{\frac{5}{96}}\,\chi_{f,1}^{(24)}
+ \sqrt{\frac{49}{1440}}\,\chi_{f,2}^{(24)}
+ \sqrt{\frac{1}{480}}\,\chi_{f,1}^{(25)} \\
& - \sqrt{\frac{5}{288}}\,\chi_{f,2}^{(25)}
+ \sqrt{\frac{3}{160}}\,\chi_{f,1}^{(26)}
- \sqrt{\frac{5}{32}}\,\chi_{f,2}^{(26)} \\
& + \sqrt{\frac{1}{30}}\,\chi_{f,1}^{(34)}
+ \sqrt{\frac{1}{90}}\,\chi_{f,2}^{(34)}
+ \sqrt{\frac{1}{480}}\,\chi_{f,1}^{(35)} \\
& + \sqrt{\frac{1}{1440}}\,\chi_{f,2}^{(35)}
+ \sqrt{\frac{3}{160}}\,\chi_{f,1}^{(36)}
+ \sqrt{\frac{1}{160}}\,\chi_{f,2}^{(36)} \\
& + \sqrt{\frac{3}{160}}\,\chi_{f,1}^{(45)}
+ \sqrt{\frac{1}{160}}\,\chi_{f,2}^{(45)}
+ \sqrt{\frac{27}{160}}\,\chi_{f,1}^{(46)} \\
& + \sqrt{\frac{9}{160}}\,\chi_{f,2}^{(46)}
\end{aligned}
\end{equation}

\begin{equation}
\begin{aligned}
\tau_f^{(7)} \;=\;& 
+ \sqrt{\frac{1}{160}}\,\chi_{f,1}^{(13)}
+ \sqrt{\frac{5}{96}}\,\chi_{f,2}^{(13)}
+ \sqrt{\frac{1}{160}}\,\chi_{f,1}^{(14)} \\
& + \sqrt{\frac{5}{96}}\,\chi_{f,2}^{(14)}
+ \sqrt{\frac{1}{160}}\,\chi_{f,1}^{(15)}
- \sqrt{\frac{1}{480}}\,\chi_{f,2}^{(15)} \\
& + \sqrt{\frac{9}{160}}\,\chi_{f,1}^{(16)}
- \sqrt{\frac{3}{160}}\,\chi_{f,2}^{(16)}
- \sqrt{\frac{1}{160}}\,\chi_{f,1}^{(23)} \\
& - \sqrt{\frac{5}{96}}\,\chi_{f,2}^{(23)}
- \sqrt{\frac{1}{160}}\,\chi_{f,1}^{(24)}
- \sqrt{\frac{5}{96}}\,\chi_{f,2}^{(24)} \\
& - \sqrt{\frac{1}{160}}\,\chi_{f,1}^{(25)}
+ \sqrt{\frac{1}{480}}\,\chi_{f,2}^{(25)}
- \sqrt{\frac{9}{160}}\,\chi_{f,1}^{(26)} \\
& + \sqrt{\frac{3}{160}}\,\chi_{f,2}^{(26)}
+ \sqrt{\frac{1}{40}}\,\chi_{f,1}^{(34)}
- \sqrt{\frac{3}{40}}\,\chi_{f,2}^{(34)} \\
& - \sqrt{\frac{1}{160}}\,\chi_{f,1}^{(35)}
+ \sqrt{\frac{3}{160}}\,\chi_{f,2}^{(35)}
- \sqrt{\frac{9}{160}}\,\chi_{f,1}^{(36)} \\
& + \sqrt{\frac{27}{160}}\,\chi_{f,2}^{(36)}
- \sqrt{\frac{1}{160}}\,\chi_{f,1}^{(45)}
+ \sqrt{\frac{3}{160}}\,\chi_{f,2}^{(45)} \\
& - \sqrt{\frac{9}{160}}\,\chi_{f,1}^{(46)}
+ \sqrt{\frac{27}{160}}\,\chi_{f,2}^{(46)}
\end{aligned}
\end{equation}

\begin{equation}
\begin{aligned}
\tau_f^{(8)} \;=\;& 
- \sqrt{\frac{1}{40}}\,\chi_{f,1}^{(12)}
- \sqrt{\frac{1}{120}}\,\chi_{f,2}^{(12)}
+ \sqrt{\frac{1}{160}}\,\chi_{f,1}^{(13)} \\
& + \sqrt{\frac{1}{480}}\,\chi_{f,2}^{(13)}
+ \sqrt{\frac{1}{160}}\,\chi_{f,1}^{(14)}
+ \sqrt{\frac{1}{480}}\,\chi_{f,2}^{(14)} \\
& + \sqrt{\frac{9}{160}}\,\chi_{f,1}^{(15)}
- \sqrt{\frac{3}{160}}\,\chi_{f,2}^{(15)}
+ \sqrt{\frac{9}{160}}\,\chi_{f,1}^{(16)} \\
& - \sqrt{\frac{3}{160}}\,\chi_{f,2}^{(16)}
+ \sqrt{\frac{1}{160}}\,\chi_{f,1}^{(23)}
+ \sqrt{\frac{1}{480}}\,\chi_{f,2}^{(23)} \\
& + \sqrt{\frac{1}{160}}\,\chi_{f,1}^{(24)}
+ \sqrt{\frac{1}{480}}\,\chi_{f,2}^{(24)}
+ \sqrt{\frac{9}{160}}\,\chi_{f,1}^{(25)} \\
& - \sqrt{\frac{3}{160}}\,\chi_{f,2}^{(25)}
+ \sqrt{\frac{9}{160}}\,\chi_{f,1}^{(26)}
- \sqrt{\frac{3}{160}}\,\chi_{f,2}^{(26)} \\
& - \sqrt{\frac{1}{40}}\,\chi_{f,1}^{(34)}
- \sqrt{\frac{1}{120}}\,\chi_{f,2}^{(34)}
- \sqrt{\frac{9}{160}}\,\chi_{f,1}^{(35)} \\
& - \sqrt{\frac{3}{160}}\,\chi_{f,2}^{(35)}
- \sqrt{\frac{9}{160}}\,\chi_{f,1}^{(36)}
- \sqrt{\frac{3}{160}}\,\chi_{f,2}^{(36)} \\
& - \sqrt{\frac{9}{160}}\,\chi_{f,1}^{(45)}
- \sqrt{\frac{3}{160}}\,\chi_{f,2}^{(45)}
- \sqrt{\frac{9}{160}}\,\chi_{f,1}^{(46)} \\
& - \sqrt{\frac{3}{160}}\,\chi_{f,2}^{(46)}
- \sqrt{\frac{9}{40}}\,\chi_{f,1}^{(56)}
- \sqrt{\frac{3}{40}}\,\chi_{f,2}^{(56)}
\end{aligned}
\end{equation}

\begin{equation}
\begin{aligned}
\tau_f^{(9)} \;=\;& 
+ \sqrt{\frac{3}{160}}\,\chi_{f,1}^{(13)}
+ \sqrt{\frac{1}{160}}\,\chi_{f,2}^{(13)}
- \sqrt{\frac{3}{160}}\,\chi_{f,1}^{(14)} \\
& - \sqrt{\frac{1}{160}}\,\chi_{f,2}^{(14)}
- \sqrt{\frac{3}{160}}\,\chi_{f,1}^{(15)}
- \sqrt{\frac{9}{160}}\,\chi_{f,2}^{(15)} \\
& - \sqrt{\frac{3}{160}}\,\chi_{f,1}^{(16)}
- \sqrt{\frac{9}{160}}\,\chi_{f,2}^{(16)}
- \sqrt{\frac{3}{160}}\,\chi_{f,1}^{(23)} \\
& - \sqrt{\frac{1}{160}}\,\chi_{f,2}^{(23)}
+ \sqrt{\frac{3}{160}}\,\chi_{f,1}^{(24)}
+ \sqrt{\frac{1}{160}}\,\chi_{f,2}^{(24)} \\
& + \sqrt{\frac{3}{160}}\,\chi_{f,1}^{(25)}
+ \sqrt{\frac{9}{160}}\,\chi_{f,2}^{(25)}
+ \sqrt{\frac{3}{160}}\,\chi_{f,1}^{(26)} \\
& + \sqrt{\frac{9}{160}}\,\chi_{f,2}^{(26)}
- \sqrt{\frac{3}{160}}\,\chi_{f,1}^{(35)}
+ \sqrt{\frac{9}{160}}\,\chi_{f,2}^{(35)} \\
& - \sqrt{\frac{3}{160}}\,\chi_{f,1}^{(36)}
+ \sqrt{\frac{9}{160}}\,\chi_{f,2}^{(36)}
+ \sqrt{\frac{3}{160}}\,\chi_{f,1}^{(45)} \\
& - \sqrt{\frac{9}{160}}\,\chi_{f,2}^{(45)}
+ \sqrt{\frac{3}{160}}\,\chi_{f,1}^{(46)}
- \sqrt{\frac{9}{160}}\,\chi_{f,2}^{(46)} \\
& + \sqrt{\frac{3}{40}}\,\chi_{f,1}^{(56)}
- \sqrt{\frac{9}{40}}\,\chi_{f,2}^{(56)}
\end{aligned}
\end{equation}

\section{Apendix II}

The wave functions corresponding to $S$= 0 are:
\begin{align}
\Phi_0^{(1)} \;=\;& \frac{1}{2\sqrt{2}}\Big(
|\uparrow\downarrow\uparrow\downarrow\uparrow\downarrow\rangle
-|\uparrow\downarrow\uparrow\downarrow\downarrow\uparrow\rangle \nonumber\\
&-|\uparrow\downarrow\downarrow\uparrow\uparrow\downarrow\rangle
+|\uparrow\downarrow\downarrow\uparrow\downarrow\uparrow\rangle \nonumber\\
&-|\downarrow\uparrow\uparrow\downarrow\uparrow\downarrow\rangle
+|\downarrow\uparrow\uparrow\downarrow\downarrow\uparrow\rangle \nonumber\\
&+|\downarrow\uparrow\downarrow\uparrow\uparrow\downarrow\rangle
-|\downarrow\uparrow\downarrow\uparrow\downarrow\uparrow\rangle
\Big)\,,
\end{align}

\begin{align}
\Phi_0^{(2)} \;=\;& \frac{1}{\sqrt{6}}\Big(
|\uparrow\downarrow\uparrow\uparrow\downarrow\downarrow\rangle
+|\uparrow\downarrow\downarrow\downarrow\uparrow\uparrow\rangle \nonumber\\
&-|\downarrow\uparrow\uparrow\uparrow\downarrow\downarrow\rangle
-|\downarrow\uparrow\downarrow\downarrow\uparrow\uparrow\rangle
\Big) \nonumber\\
&-\frac{1}{2\sqrt{6}}\Big(
|\uparrow\downarrow\uparrow\downarrow\uparrow\downarrow\rangle
+|\uparrow\downarrow\uparrow\downarrow\downarrow\uparrow\rangle \nonumber\\
&+|\uparrow\downarrow\downarrow\uparrow\uparrow\downarrow\rangle
+|\uparrow\downarrow\downarrow\uparrow\downarrow\uparrow\rangle \nonumber\\
&-|\downarrow\uparrow\uparrow\downarrow\uparrow\downarrow\rangle
-|\downarrow\uparrow\uparrow\downarrow\downarrow\uparrow\rangle \nonumber\\
&-|\downarrow\uparrow\downarrow\uparrow\uparrow\downarrow\rangle
-|\downarrow\uparrow\downarrow\uparrow\downarrow\uparrow\rangle
\Big)\,,
\end{align}

\begin{align}
\Phi_0^{(3)} \;=\;& \frac{1}{\sqrt{6}}\Big(
|\uparrow\uparrow\downarrow\downarrow\uparrow\downarrow\rangle
-|\uparrow\uparrow\downarrow\downarrow\downarrow\uparrow\rangle \nonumber\\
&+|\downarrow\downarrow\uparrow\uparrow\uparrow\downarrow\rangle
-|\downarrow\downarrow\uparrow\uparrow\downarrow\uparrow\rangle
\Big) \nonumber\\
&-\frac{1}{2\sqrt{6}}\Big(
|\uparrow\downarrow\uparrow\downarrow\uparrow\downarrow\rangle
-|\uparrow\downarrow\uparrow\downarrow\downarrow\uparrow\rangle \nonumber\\
&+|\uparrow\downarrow\downarrow\uparrow\uparrow\downarrow\rangle
-|\uparrow\downarrow\downarrow\uparrow\downarrow\uparrow\rangle \nonumber\\
&+|\downarrow\uparrow\uparrow\downarrow\uparrow\downarrow\rangle
-|\downarrow\uparrow\uparrow\downarrow\downarrow\uparrow\rangle \nonumber\\
&+|\downarrow\uparrow\downarrow\uparrow\uparrow\downarrow\rangle
-|\downarrow\uparrow\downarrow\uparrow\downarrow\uparrow\rangle
\Big)\,,
\end{align}

\begin{align}
\Phi_0^{(4)} \;=\;& \frac{\sqrt{2}}{3}\Big(
|\uparrow\uparrow\downarrow\uparrow\downarrow\downarrow\rangle
-|\downarrow\downarrow\uparrow\downarrow\uparrow\uparrow\rangle
\Big) \nonumber\\
&-\frac{\sqrt{2}}{6}\Big(
|\uparrow\uparrow\downarrow\downarrow\uparrow\downarrow\rangle
+|\uparrow\uparrow\downarrow\downarrow\downarrow\uparrow\rangle \nonumber\\
&+|\uparrow\downarrow\uparrow\uparrow\downarrow\downarrow\rangle
+|\downarrow\uparrow\uparrow\uparrow\downarrow\downarrow\rangle
\Big) \nonumber\\
&+\frac{\sqrt{2}}{6}\Big(
|\uparrow\downarrow\downarrow\downarrow\uparrow\uparrow\rangle
+|\downarrow\uparrow\downarrow\downarrow\uparrow\uparrow\rangle \nonumber\\
&+|\downarrow\downarrow\uparrow\uparrow\uparrow\downarrow\rangle
+|\downarrow\downarrow\uparrow\uparrow\downarrow\uparrow\rangle
\Big) \nonumber\\
&+\frac{\sqrt{2}}{12}\Big(
|\uparrow\downarrow\uparrow\downarrow\uparrow\downarrow\rangle
+|\uparrow\downarrow\uparrow\downarrow\downarrow\uparrow\rangle \nonumber\\
&-|\uparrow\downarrow\downarrow\uparrow\uparrow\downarrow\rangle
-|\uparrow\downarrow\downarrow\uparrow\downarrow\uparrow\rangle \nonumber\\
&+|\downarrow\uparrow\uparrow\downarrow\uparrow\downarrow\rangle
+|\downarrow\uparrow\uparrow\downarrow\downarrow\uparrow\rangle \nonumber\\
&-|\downarrow\uparrow\downarrow\uparrow\uparrow\downarrow\rangle
-|\downarrow\uparrow\downarrow\uparrow\downarrow\uparrow\rangle
\Big)\,,
\end{align}

\begin{align}
\Phi_0^{(5)} \;=\;& \frac{1}{2}\Big(
|\uparrow\uparrow\uparrow\downarrow\downarrow\downarrow\rangle
-|\downarrow\downarrow\downarrow\uparrow\uparrow\uparrow\rangle
\Big) \nonumber\\
&-\frac{1}{6}\Big(
|\uparrow\uparrow\downarrow\uparrow\downarrow\downarrow\rangle
+|\uparrow\uparrow\downarrow\downarrow\uparrow\downarrow\rangle \nonumber\\
&+|\uparrow\uparrow\downarrow\downarrow\downarrow\uparrow\rangle
+|\uparrow\downarrow\uparrow\uparrow\downarrow\downarrow\rangle \nonumber\\
&+|\uparrow\downarrow\uparrow\downarrow\uparrow\downarrow\rangle
+|\uparrow\downarrow\uparrow\downarrow\downarrow\uparrow\rangle \nonumber\\
&+|\downarrow\uparrow\uparrow\uparrow\downarrow\downarrow\rangle
+|\downarrow\uparrow\uparrow\downarrow\uparrow\downarrow\rangle \nonumber\\
&+|\downarrow\uparrow\uparrow\downarrow\downarrow\uparrow\rangle
\Big) \nonumber\\
&+\frac{1}{6}\Big(
|\uparrow\downarrow\downarrow\uparrow\uparrow\downarrow\rangle
+|\uparrow\downarrow\downarrow\uparrow\downarrow\uparrow\rangle \nonumber\\
&+|\uparrow\downarrow\downarrow\downarrow\uparrow\uparrow\rangle
+|\downarrow\uparrow\downarrow\uparrow\uparrow\downarrow\rangle \nonumber\\
&+|\downarrow\uparrow\downarrow\uparrow\downarrow\uparrow\rangle
+|\downarrow\uparrow\downarrow\downarrow\uparrow\uparrow\rangle \nonumber\\
&+|\downarrow\downarrow\uparrow\uparrow\uparrow\downarrow\rangle
+|\downarrow\downarrow\uparrow\uparrow\downarrow\uparrow\rangle \nonumber\\
&+|\downarrow\downarrow\uparrow\downarrow\uparrow\uparrow\rangle
\Big)\,.
\end{align}

These are  9 spin functions with $S$=1:
\begin{align}
\Phi_1^{(1)} \;=\;& \frac{1}{2}\Big(
|\uparrow\uparrow\uparrow\downarrow\uparrow\downarrow\rangle
-|\uparrow\uparrow\uparrow\downarrow\downarrow\uparrow\rangle \nonumber\\
&-|\uparrow\uparrow\downarrow\uparrow\uparrow\downarrow\rangle
+|\uparrow\uparrow\downarrow\uparrow\downarrow\uparrow\rangle
\Big)\,,
\end{align}

\begin{align}
\Phi_1^{(2)} \;=\;& \frac{1}{2}\Big(
|\uparrow\uparrow\uparrow\uparrow\downarrow\downarrow\rangle
-|\uparrow\uparrow\uparrow\downarrow\downarrow\uparrow\rangle \nonumber\\
&-|\uparrow\uparrow\downarrow\uparrow\uparrow\downarrow\rangle
+|\uparrow\uparrow\downarrow\downarrow\uparrow\uparrow\rangle
\Big)\,,
\end{align}

\begin{align}
\Phi_1^{(3)} \;=\;& \frac{1}{2}\Big(
|\uparrow\uparrow\uparrow\downarrow\uparrow\downarrow\rangle
-|\uparrow\uparrow\uparrow\downarrow\downarrow\uparrow\rangle \nonumber\\
&-|\uparrow\downarrow\uparrow\uparrow\uparrow\downarrow\rangle
+|\uparrow\downarrow\uparrow\uparrow\downarrow\uparrow\rangle
\Big)\,,
\end{align}

\begin{align}
\Phi_1^{(4)} \;=\;& \frac{1}{2}\Big(
|\uparrow\uparrow\uparrow\uparrow\downarrow\downarrow\rangle
-|\uparrow\uparrow\uparrow\downarrow\downarrow\uparrow\rangle \nonumber\\
&-|\uparrow\downarrow\uparrow\uparrow\uparrow\downarrow\rangle
+|\uparrow\downarrow\uparrow\downarrow\uparrow\uparrow\rangle
\Big)\,,
\end{align}

\begin{align}
\Phi_1^{(5)} \;=\;& \frac{1}{2}\Big(
|\uparrow\uparrow\uparrow\uparrow\downarrow\downarrow\rangle
-|\uparrow\uparrow\downarrow\uparrow\downarrow\uparrow\rangle \nonumber\\
&-|\uparrow\downarrow\uparrow\uparrow\uparrow\downarrow\rangle
+|\uparrow\downarrow\downarrow\uparrow\uparrow\uparrow\rangle
\Big)\,,
\end{align}

\begin{align}
\Phi_1^{(6)} \;=\;& \frac{1}{2}\Big(
|\uparrow\uparrow\uparrow\downarrow\uparrow\downarrow\rangle
-|\uparrow\uparrow\uparrow\downarrow\downarrow\uparrow\rangle \nonumber\\
&-|\downarrow\uparrow\uparrow\uparrow\uparrow\downarrow\rangle
+|\downarrow\uparrow\uparrow\uparrow\downarrow\uparrow\rangle
\Big)\,,
\end{align}

\begin{align}
\Phi_1^{(7)} \;=\;& \frac{1}{2}\Big(
|\uparrow\uparrow\uparrow\uparrow\downarrow\downarrow\rangle
-|\uparrow\uparrow\uparrow\downarrow\downarrow\uparrow\rangle \nonumber\\
&-|\downarrow\uparrow\uparrow\uparrow\uparrow\downarrow\rangle
+|\downarrow\uparrow\uparrow\downarrow\uparrow\uparrow\rangle
\Big)\,,
\end{align}

\begin{align}
\Phi_1^{(8)} \;=\;& \frac{1}{2}\Big(
|\uparrow\uparrow\uparrow\uparrow\downarrow\downarrow\rangle
-|\uparrow\uparrow\downarrow\uparrow\downarrow\uparrow\rangle \nonumber\\
&-|\downarrow\uparrow\uparrow\uparrow\uparrow\downarrow\rangle
+|\downarrow\uparrow\downarrow\uparrow\uparrow\uparrow\rangle
\Big)\,,
\end{align}

\begin{align}
\Phi_1^{(9)} \;=\;& \frac{1}{2}\Big(
|\uparrow\uparrow\uparrow\uparrow\downarrow\downarrow\rangle
-|\uparrow\downarrow\uparrow\uparrow\downarrow\uparrow\rangle \nonumber\\
&-|\downarrow\uparrow\uparrow\uparrow\uparrow\downarrow\rangle
+|\downarrow\downarrow\uparrow\uparrow\uparrow\uparrow\rangle
\Big)\,.
\end{align}
Below are the 5 wavefuncions for $S$=2. 

\begin{align}
\Phi_2^{(1)} \;=\;& \frac{1}{\sqrt{2}}\Big(
|\uparrow\uparrow\uparrow\uparrow\uparrow\downarrow\rangle
-|\uparrow\uparrow\uparrow\uparrow\downarrow\uparrow\rangle
\Big)\,,
\end{align}

\begin{align}
\Phi_2^{(2)} \;=\;& \frac{1}{\sqrt{6}}\Big(
|\uparrow\uparrow\uparrow\uparrow\uparrow\downarrow\rangle
+|\uparrow\uparrow\uparrow\uparrow\downarrow\uparrow\rangle \nonumber\\
&-2|\uparrow\uparrow\uparrow\downarrow\uparrow\uparrow\rangle
\Big)\,,
\end{align}

\begin{align}
\Phi_2^{(3)} \;=\;& \frac{1}{2\sqrt{3}}\Big(
|\uparrow\uparrow\uparrow\uparrow\uparrow\downarrow\rangle
+|\uparrow\uparrow\uparrow\uparrow\downarrow\uparrow\rangle \nonumber\\
&+|\uparrow\uparrow\uparrow\downarrow\uparrow\uparrow\rangle
-3|\uparrow\uparrow\downarrow\uparrow\uparrow\uparrow\rangle
\Big)\,,
\end{align}

\begin{align}
\Phi_2^{(4)} \;=\;& \frac{1}{2\sqrt{5}}\Big(
|\uparrow\uparrow\uparrow\uparrow\uparrow\downarrow\rangle
+|\uparrow\uparrow\uparrow\uparrow\downarrow\uparrow\rangle \nonumber\\
&+|\uparrow\uparrow\uparrow\downarrow\uparrow\uparrow\rangle
+|\uparrow\uparrow\downarrow\uparrow\uparrow\uparrow\rangle \nonumber\\
&-4|\uparrow\downarrow\uparrow\uparrow\uparrow\uparrow\rangle
\Big)\,,
\end{align}

\begin{align}
\Phi_2^{(5)} \;=\;& \frac{1}{\sqrt{30}}\Big(
|\uparrow\uparrow\uparrow\uparrow\uparrow\downarrow\rangle
+|\uparrow\uparrow\uparrow\uparrow\downarrow\uparrow\rangle \nonumber\\
&+|\uparrow\uparrow\uparrow\downarrow\uparrow\uparrow\rangle
+|\uparrow\uparrow\downarrow\uparrow\uparrow\uparrow\rangle \nonumber\\
&+|\uparrow\downarrow\uparrow\uparrow\uparrow\uparrow\rangle
-5|\downarrow\uparrow\uparrow\uparrow\uparrow\uparrow\rangle
\Big)\,.
\end{align}

And, finally the function for $S$=3 reads:

\begin{equation}
\Phi_3^{(1)} \;=\; |\uparrow\uparrow\uparrow\uparrow\uparrow\uparrow\rangle\,.
\end{equation}
\bibliography{hdibarion}

\end{document}